\documentclass[12pt,a4paper]{article}
\usepackage[dvips]{graphicx,color}
\newcommand{\be}{\begin{equation}}
\newcommand{\ee}{\end{equation}}
\newcommand{\ba}{\begin{eqnarray}}
\newcommand{\ea}{\end{eqnarray}}
\newcommand{\baa}{\begin{eqnarray*}}
\newcommand{\eaa}{\end{eqnarray*}}
\newcommand{\lab}[1]{\label{#1}}

\newcommand{\dis}{\displaystyle}
\newcommand{\biq}{\mbox{\boldmath $q$}}
\newcommand{\bip}{\mbox{\boldmath $p$}}
\newcommand{\bif}{\mbox{\boldmath $f$}}
\newcommand{\biL}{\mbox{\boldmath $\Lambda$}}
\newcommand{\bil}{\mbox{\boldmath $\lambda$}}

\begin{document}
{\pagestyle{empty}
\vskip 1.5cm

{\renewcommand{\thefootnote}{\fnsymbol{footnote}}
\centerline{\large \bf Generalized-Ensemble Algorithms: 
Enhanced Sampling Techniques}
\centerline{\large \bf for Monte Carlo and Molecular Dynamics
Simulations}
%\vskip 3.0cm
\vskip 1.0cm

\centerline{Yuko Okamoto\footnote{\ \ 
e-mail: okamotoy@ims.ac.jp; URL: http://konf2.ims.ac.jp/}}
%\vskip 1.5cm
\centerline{{\it Department of Theoretical Studies}}
\centerline{{\it Institute for Molecular Science}}
\centerline{{\it Okazaki, Aichi 444-8585, Japan}}
\centerline{{\rm and}}
\centerline{{\it Department of Functional Molecular Science}}
\centerline{{\it The Graduate University for Advanced Studies}}
\centerline{{\it Okazaki, Aichi 444-8585, Japan}}

%\vskip 1.5cm
\vskip 1.0cm

%\centerline{{\it {\bf Keywords:} protein folding; 
%generalized-ensemble algorithm; multicanonical algorithm;}}
%\centerline{{\it simulated tempering; replica-exchange method; parallel tempering}}

\medbreak
%\vskip 1.0cm
 
\centerline{\bf ABSTRACT}
\vskip 0.3cm

In complex systems with many degrees of freedom such as
spin glass and biomolecular systems, 
conventional simulations 
in canonical ensemble suffer from the quasi-ergodicity
problem.
A simulation in generalized ensemble performs a random walk
in potential energy space
and overcomes this difficulty.
From only one simulation run, one can
obtain canonical-ensemble averages of physical quantities as
functions of temperature
by the single-histogram and/or multiple-histogram reweighting techniques.
In this article we review the generalized-ensemble algorithms.
Three well-known methods, namely, 
multicanonical algorithm, simulated tempering, and 
replica-exchange method, are described first.
Both Monte Carlo and molecular dynamics versions of the 
algorithms are given.
We then present five new generalized-ensemble algorithms
which are extensions of the above methods.
 
%\vfill
%\newpage}
}
%\baselineskip=0.8cm
%%%%%%%%%%%%%%%%%%%%%%% Section 1 %%%%%%%%%%%%%%%%%%%%%%%%
%\noindent
%{\bf 1. INTRODUCTION} \\
\section{INTRODUCTION}

Since the pioneering work of Metropolis and coworkers \cite{Metro}
half a century ago,
computer simulations have been indispensable means of research
in many fields of physical sciences.
In the field of molecular science, for instance,  
a number of powerful simulation algorithms have been 
developed (for reviews see, e.g.,
Refs.~\cite{RevSch}--\cite{RevHO2}).

Canonical fixed-temperature 
%Despite the great advancement of
%computer and computational technologies in the past decades, 
%however, 
simulations of complex systems such as spin glasses and 
biopolymers are greatly hampered by the multiple-minima
problem, or the quasi-ergodicity problem.
Because simulations at low temperatures tend to get
trapped in one of huge number of local-minimum-energy states,
it is very difficult to obtain accurate canonical distributions
at low temperatures by conventional Monte Carlo (MC) and
molecular dynamics (MD) methods.
One way to overcome this multiple-minima
problem is to perform a simulation in a {\it generalized ensemble} 
where each state is weighted by an artificial,
non-Boltzmann probability
weight factor so that
a random walk in potential energy space may be realized
(for reviews see, e.g., Refs.~\cite{RevHO1}--\cite{Be02}).
The random walk allows the simulation to escape from any
energy barrier and to sample much wider configurational
space than by conventional methods.
Monitoring the energy in a single simulation run, one can
obtain not only
the global-minimum-energy state but also canonical ensemble
averages as functions of temperature by the single-histogram \cite{FS1}
and/or multiple-histogram \cite{FS2,WHAM} reweighting techniques
(an extension of the multiple-histogram method is also referred to as
{\it weighted histogram analysis method} (WHAM) \cite{WHAM}).
Besides generalized-ensemble algorithms, which are usually
based on local updates, methods based on non-local updates
such as cluster algorithms and their generalizations have
also been widely used \cite{SW87}--\cite{ELM93}.
In this article, we focus our discussion on generalized-ensemble
algorithms.

One of the most well-known generalized-ensemble methods is perhaps
{\it multicanonical algorithm} (MUCA) \cite{MUCA1,MUCA2}
%(for reviews see, e.g., Refs.~\cite{MUCArev,RevJanke}).
(for a review see, e.g., Ref.~\cite{MUCArev}).
(The method is also referred to as {\it entropic sampling}
\cite{Lee}, {\it adaptive umbrella sampling} \cite{MZ} 
{\it of the potential energy} \cite{BK},
{\it random walk algorithm} \cite{Landau1,Landau2},
and {\it density of states Monte Carlo} \cite{dePablo}.
MUCA can also be considered as a sophisticated, ideal realization of
a class of algorithms called {\it umbrella sampling}
\cite{US}.  Also closely related methods are {\it transition
matrix methods} reviewed in Refs.~\cite{WS02,Be02}.)
%The mathematical equivalence of multicanonical algorithm and entropic
%sampling has been given in Ref.~\cite{BHO}.)
MUCA and its generalizations have been applied to
%spin glass systems
spin systems
%(see, e.g., Refs.~\cite{MUCA3}--\cite{MUCA6}).
(see, e.g., Refs.~\cite{MUCA3}--\cite{BBJ00}).
MUCA was also introduced to the molecular simulation field
\cite{HO}.
%(for previous reviews of generalized-ensemble approach in the
%protein folding problem, see, e.g.,
%Refs.~\cite{RevO}--\cite{RevHO2}).
Since then MUCA and its generalizations have been extensively
used in many applications in
protein and related systems \cite{HO94}--\cite{OO03}.
Molecular dynamics version of MUCA
has also been developed \cite{HOE96,NNK,BK} (see also
Refs.~\cite{Muna,HOE96} for Langevin dynamics version).
MUCA has been extended so that flat distributions in
other parameters instead of potential energy may be 
obtained \cite{BeHaNe93,JK95,KPV,BK2,ICK,BNO03}.
Moreover, multidimensional (or multicomponent) extensions of MUCA
can be found in Refs.~\cite{KPV,BK2,HNSKN,OO03}.
  
While a simulation in multicanonical ensemble performs a free
1D random walk in potential energy space, that in
{\it simulated tempering} (ST) \cite{ST1,ST2} 
(the method is also referred to as the
{\it method of expanded ensemble} \cite{ST1})
performs a free random walk in temperature space
(for a review, see, e.g., Ref.~\cite{STrev}).
This random walk, in turn,
induces a random walk in potential energy space and
allows the simulation to escape from
states of energy local minima.
ST has also been applied to protein folding
problem \cite{IRB1,HO96a,HO96b,IRB2}.

%A third generalized-ensemble algorithm that is related
%to MUCA is {\it 1/k-sampling} \cite{HS}.
%A simulation in 1/k-sampling
%performs a free random walk in 
%entropy space, which, in turn,
%induces a random walk in potential energy space.
%The relation among the above three generalized-ensemble
%algorithms was discussed and the effectiveness of the
%three methods in protein
%folding problem was compared
%\cite{HO96b}.  

The generalized-ensemble method
is powerful, but in the above two methods the probability
weight factors are not {\it a priori} known and have to be
determined by iterations of short trial simulations.
This process can be non-trivial and very tedius for
complex systems with many degreees of freedom.
%local-minimum-energy states.
Therefore, there have been attempts to accelerate
the convergence of the iterative process for MUCA
weight factor determination
\cite{MUCA3,KPV,SmBr,H97c,MUCAW,BK}
(see also Refs.~\cite{MUCArev,Janke03}).
  
%A new generalized-ensemble algorithm that is based on
%the weight factor of Tsallis statistical mechanics \cite{Tsa}
%was recently developed with the hope of overcoming this difficulty
%\cite{HO96d,HEO98},
%and the method was applied to a peptide folding problem
%\cite{HMO97,HOO}.  A similar but slightly different formulation 
%is given in Ref.~\cite{Str2}.  See also Ref.~\cite{Muna2} for
%a combination of Tsallis statistics with simulated tempering.
%(Optimization problems were also addressed by 
%simulated annealing 
%algorithms \cite{SA} based on the Tsallis weight in
%Refs.~\cite{STsal}--\cite{H97b}.  For reviews of molecular simulations
%based on Tsallis statistics, see, e.g., 
%Refs.~\cite{BeSt}--\cite{RevHO3}.)
%In this generalized ensemble the weight factor is known,
%once the value of the global-minimum energy is given \cite{HO96d}.
%The advantage of this ensemble is that it greatly simplifies the
%determination of the weight factor.
%However, the estimation of
%the global-minimum energy can still be very difficult.

In the {\it replica-exchange method} (REM)
\cite{RE1}--\cite{RE2}, the difficulty of weight factor
determination is greatly alleviated.  (A closely related method 
was independently developed in Ref.~\cite{RE3}.
Similar methods in which the same equations are used but
emphasis is laid on optimizations have
been developed \cite{KT,JWK}. 
REM is also referred to as 
{\it multiple Markov chain method} \cite{RE4}
and {\it parallel tempering} \cite{STrev}.  Details
of literature
about REM and related algorithms can be found in
recent reviews \cite{IBArev,RevMSO}.)
In this method, a number of
non-interacting copies (or replicas) of the original system
at different temperatures are
simulated independently and
simultaneously by the conventional MC or MD method. Every few steps,
pairs of replicas are exchanged with a specified transition
probability.
The weight factor is just the
product of Boltzmann factors, and so it is essentially known.

REM has already been used in many applications in protein 
systems \cite{H97,SO,IRB2}\cite{WD}--\cite{LMMO03}. 
Other molecular simulation fields have also been studied by
this method in various ensembles \cite{FD}--\cite{OKOM}.
Moreover, REM was applied to cluster studies in quantum
chemistry field \cite{ISNO}.
The details of molecular dynamics algorithm have been worked out
for REM in Ref.~\cite{SO} (see also Refs.~\cite{H97,Yama}).
This led to a wide application of replica-exchange molecular
dynamics method in the protein folding problem \cite{Gar}-\cite{SRNP03}.
%We then developed a multidimensional REM which is particularly
%useful in free energy calculations \cite{SKO}
%(see also Refs.~\cite{Huk2,YP}).

However, REM also has a computational difficulty:
As the number of degrees of freedom of the system increases,
the required number of replicas also greatly increases, whereas 
only a single replica is simulated in MUCA or ST.
This demands a lot of computer power for complex systems.
Our solution to this problem is: Use REM for the weight
factor determinations of MUCA or ST, which is much
simpler than previous iterative methods of weight
determinations, and then perform a long MUCA or ST
production run.
The first example is
the {\it replica-exchange multicanonical algorithm} (REMUCA)
\cite{SO3,MSO03}.
In REMUCA,
a short replica-exchange simulation is performed, and the 
multicanonical weight factor is determined by
the multiple-histogram reweighting techniques \cite{FS2,WHAM}.
Another example of such a combination is the
{\it replica-exchange simulated tempering} (REST) \cite{MO4}.
In REST, a short replica-exchange simulation is performed, and
the simulated tempering weight factor is determined by
the multiple-histogram reweighting techniques \cite{FS2,WHAM}.

We have introduced two further extensions of REM,
which we refer to as {\it multicanonical replica-exchange method}
(MUCAREM) \cite{SO3,MSO03} (see also Refs.~\cite{XB00,FYD02}) 
and {\it simulated tempering replica-exchange method}
(STREM) \cite{STREM}.
In MUCAREM, a replica-exchange simulation is
performed
with a small number of replicas each in multicanonical ensemble
of different energy ranges.  In STREM, on the other hand,
a replica-exchange simulation is performed with a small 
number of replicas in ``simulated tempering'' ensemble
of different temperature ranges.

Finally, one is naturally led to a multidimensional 
(or, multivariable) extension of REM, which we refer
to as {\it multidimensional replica-exhcange method} (MREM)
\cite{SKO} (see also Refs.~\cite{Huk2,YP,WBS02,FE03}).  A special 
realization 
of MREM is {\it replica-exchange umbrella sampling} (REUS)
\cite{SKO} and it is particularly useful in free energy calculations.

In this article, we describe the eight generalized-ensemble algorithms
mentioned above.  Namely, we first review three familiar methods:
MUCA, ST, and REM.  We then present the five new algorithms:
REMUCA, REST, MUCAREM, STREM, and MREM (and REUS). 

%\noindent
%{\bf 2. METHODS}\\
\section{GENERALIZED-ENSEMBLE ALGORITHMS}

\subsection{Multicanonical Algorithm and Simulated Tempering}

Let us consider a system of $N$ atoms of 
mass $m_k$ ($k=1, \cdots, N$)
with their coordinate vectors and
momentum vectors denoted by 
$q \equiv \{{\biq}_1, \cdots, {\biq}_N\}$ and 
$p \equiv \{{\bip}_1, \cdots, {\bip}_N\}$,
respectively.
The Hamiltonian $H(q,p)$ of the system is the sum of the
kinetic energy $K(p)$ and the potential energy $E(q)$:
\begin{equation}
H(q,p) =  K(p) + E(q)~,
\label{eqn1}
\end{equation}
where
\begin{equation}
K(p) =  \sum_{k=1}^N \frac{{\bip_k}^2}{2 m_k}~.
\label{eqn2}
\end{equation}

In the canonical ensemble at temperature $T$ 
each state $x \equiv (q,p)$ with the Hamiltonian $H(q,p)$
is weighted by the Boltzmann factor:
\begin{equation}
%W_{\rm B}(x;T) = e^{-\beta H(q,p)}~,
W_{\rm B}(x;T) = \exp \left(-\beta H(q,p) \right)~,
\label{eqn3}
\end{equation}
where the inverse temperature $\beta$ is defined by 
$\beta = 1/k_{\rm B} T$ ($k_{\rm B}$ is the Boltzmann constant). 
The average kinetic energy at temperature $T$ is then given by
\begin{equation}
\left< ~K(p)~ \right>_T =  
\left< \sum_{k=1}^N \frac{{\bip_k}^2}{2 m_k} \right>_T 
= \frac{3}{2} N k_{\rm B} T~.
\label{eqn4}
\end{equation}

Because the coordinates $q$ and momenta $p$ are decoupled
in Eq.~(\ref{eqn1}), we can suppress the kinetic energy
part and can write the Boltzmann factor as
\begin{equation}
%W_{\rm B}(x;T) = W_{\rm B}(E;T) = e^{-\beta E}~.
W_{\rm B}(x;T) = W_{\rm B}(E;T) = \exp (-\beta E)~.
\label{eqn4b}
\end{equation}
The canonical probability distribution of potential energy
$P_{\rm B}(E;T)$ is then given by the product of the
density of states $n(E)$ and the Boltzmann weight factor
$W_{\rm B}(E;T)$:
\begin{equation}
 P_{\rm B}(E;T) \propto n(E) W_{\rm B}(E;T)~.
\label{eqn4c}
\end{equation}
Since $n(E)$ is a rapidly
increasing function and the Boltzmann factor decreases exponentially,
the canonical ensemble yields a bell-shaped distribution
which has a maximum around the average energy at temperature $T$. 
The
conventional MC or MD simulations at constant temperature
are expected to yield $P_{\rm B}(E;T)$.
A MC simulation based on the Metropolis 
algorithm \cite{Metro} is performed with the following 
transition probability from a state $x$
of potential energy $E$ to a state $x^{\prime}$ of 
potential energy $E^{\prime}$:
\begin{equation}
w(x \rightarrow x^{\prime})
= {\rm min} \left(1,\frac{W_{\rm B}(E^{\prime};T)}{W_{\rm B}(E;T)}\right)
%= {\rm min} \left(1,e^{-\beta (E^{\prime} - E)}\right)~.
= {\rm min} \left(1,\exp \left[-\beta (E^{\prime} - E) \right]\right)~.
\label{eqn4d}
\end{equation}
A MD simulation, on the other hand, is based on the 
following Newton equation:
\begin{equation}
\dot{{\bip}}_k ~=~ - \frac{\partial E}{\partial {\biq}_k}
~=~ {\bif}_k~,
\label{eqn4e}
\end{equation}
where ${\bif}_k$ is the force acting on the $k$-th atom
($k = 1, \cdots, N$).
This equation actually yields the microcanonical ensemble, and
we have to add a thermostat such as Nos{\'e}-Hoover
algorithm \cite{Nose,Hoover} and the
constraint method \cite{HLM,EM}
in order to obtain the canonical ensemble.
However, in practice, it is very difficult to obtain accurate canonical
distributions of complex systems at low temperatures by
conventional MC or MD simulation methods.
This is because simulations at low temperatures tend to get
trapped in one or a few of local-minimum-energy states.

In the multicanonical ensemble \cite{MUCA1,MUCA2}, on the other
hand, each state is weighted by
a non-Boltzmann weight
factor $W_{\rm mu}(E)$ (which we refer to as the {\it multicanonical
weight factor}) so that a uniform potential energy
distribution $P_{\rm mu}(E)$ is obtained:
\begin{equation}
 P_{\rm mu}(E) \propto n(E) W_{\rm mu}(E) \equiv {\rm constant}~.
\label{eqn5}
\end{equation}
The flat distribution implies that
a free random walk in the potential energy space is realized
in this ensemble.
This allows the simulation to escape from any local minimum-energy states
and to sample the configurational space much more widely than 
the conventional canonical MC or MD methods.

The definition in Eq.~(\ref{eqn5}) implies that
the multicanonical
weight factor is inversely proportional to the density of
states, and we can write it as follows:
\begin{equation}
%W_{\rm mu}(E) \equiv e^{-\beta_0 E_{\rm mu}(E;T_0)}
W_{\rm mu}(E) \equiv \exp \left[-\beta_0 E_{\rm mu}(E;T_0) \right]
= \frac{1}{n(E)}~,
\label{eqn6}
\end{equation}
where we have chosen an arbitrary reference
temperature, $T_0 = 1/k_{\rm B} \beta_0$, and
the ``{\it multicanonical potential energy}''
is defined by
\begin{equation}
 E_{\rm mu}(E;T_0) \equiv k_{\rm B} T_0 \ln n(E) = T_0 S(E)~.
\label{eqn7}
\end{equation}
Here, $S(E)$ is the entropy in the microcanonical
ensemble.
Since the density of states of the system is usually unknown,
the multicanonical weight factor
has to be determined numerically by iterations of short preliminary
runs \cite{MUCA1,MUCA2}. 

A multicanonical Monte Carlo simulation
is performed, for instance, with the usual Metropolis criterion \cite{Metro}:
The transition probability of state $x$ with potential energy
$E$ to state $x^{\prime}$ with potential energy $E^{\prime}$ is given by
\begin{equation}
w(x \rightarrow x^{\prime})
= {\rm min} \left(1,\frac{W_{\rm mu}(E^{\prime})}{W_{\rm mu}(E)}\right)
= {\rm min} \left(1,\frac{n(E)}{n(E^{\prime})}\right)
= {\rm min} \left(1,\exp \left( - \beta_0 \Delta E_{\rm mu} \right)\right)~,
\label{eqn8}
\end{equation}
where
\begin{equation}
\Delta E_{\rm mu} = E_{\rm mu}(E^{\prime};T_0) - E_{\rm mu}(E;T_0)~.
\label{eqn9}
\end{equation}
The molecular dynamics algorithm in multicanonical ensemble
also naturally follows from Eq.~(\ref{eqn6}), in which the
regular constant temperature molecular dynamics simulation
(with $T=T_0$) is performed by solving the following modified
%Newton equation instead of Eq. (\ref{eqn4e}): \cite{HOE96,NNK,BK}
Newton equation instead of Eq. (\ref{eqn4e}): \cite{HOE96,NNK}  
\begin{equation}
\dot{{\bip}}_k ~=~ - \frac{\partial E_{\rm mu}(E;T_0)}{\partial {\biq}_k}
~=~ \frac{\partial E_{\rm mu}(E;T_0)}{\partial E}~{\bif}_k~.
\label{eqn9a}
\end{equation}
%with, for example, the Nos{\'e}-Hoover thermostat \cite{Nose,Hoover}.
%where ${\bif}_k$ is the usual force acting on the $k$-th atom
%($k = 1, \cdots, N$).
From Eq.~(\ref{eqn7}) this equation can be rewritten as
\begin{equation}
\dot{{\bip}}_k ~=~ \frac{T_0}{T(E)}~{\bif}_k~,
\label{eqn9b}
\end{equation}
where the following thermodynamic relation gives
the definition of the ``effective temperature''
$T(E)$:
\begin{equation}
\left. \frac{\partial S(E)}{\partial E}\right|_{E=E_a}
~=~ \frac{1}{T(E_a)}~,
\label{eqn9c}
\end{equation}
with
\begin{equation}
E_a ~=~ <E>_{T(E_a)}~.
\label{eqn9d}
\end{equation}

If the exact multicanonical weight factor $W_{\rm mu}(E)$ is known, 
one can calculate the ensemble averages of any physical quantity $A$ 
at any temperature 
$T$ ($= 1/k_{\rm B} \beta$) as follows: 
\begin{equation} 
<A>_T =  \frac{ \displaystyle{ 
\sum_E~ A (E) P_{\rm B}(E;T)} }
{\displaystyle{ \sum_E~ P_{\rm B}(E;T)} } =  \frac{ \displaystyle{ 
\sum_E~ A (E) n(E) \exp(-\beta E) } }
{\displaystyle{ \sum_E~ n(E) \exp(-\beta E) } }~, 
\label{eqn10a}
\end{equation}
where the density of states is given by (see Eq. (\ref{eqn6}))
\begin{equation} 
%n(E)= W_{\rm mu}^{-1} (E)~.
n(E)= \frac{1}{W_{\rm mu}(E)}~.
\label{eqn10b}
\end{equation}
The summation instead of integration is used
in Eq.~(\ref{eqn10a}), because we often
discretize the potential energy $E$ with step size $\epsilon$
($E = E_i; i=1, 2, \cdots$).
% and express the density of states as $n(E_i)$.
Here, the explicit form of the physical quantity $A$
should be known as a function of potential energy $E$.
For instance, $A(E)=E$ gives the average
potential energy $<E>_T$ as a function of temperature, and 
$A(E) = \beta^2 (E - <E>_T)^2$ gives specific heat.

In general, the multicanonical weight factor $W_{\rm mu}(E)$, or
the density of states $n(E)$, is not 
$a \ priori$ known, and one needs its estimator 
for a numerical simulation. 
This estimator is usually obtained 
from iterations of short trial multicanonical simulations. 
The details of this process
are described, for instance, in Refs.~\cite{MUCA3,OH}.
%(see, for instance, Refs.  
%%\cite{BC,REV4,SB,H,BMU}).
%\cite{BMU,REVMU1,REV4}).
%\cite{MU3,OH,BMU} ). 
%Once the estimator of the weight factor $W_{\rm mu}(E)$ is 
%thus determined, 
%one can, in principle, obtain the ensemble averages of
%physical quantity $A$ as a function of temperature by
%Eqs. (\ref{eqn10a}) and (\ref{eqn10b}).
%
However, the iterative process can be non-trivial and very tedius for
complex systems.
%, and there have been attempts to accelerate
%the convergence of the iterative process
%\cite{MUCA3,KPV,SmBr,H97c,MUCAW,BK,Landau1}.

In practice, it is impossible to obtain the ideal
multicanonical weight factor with completely uniform
potential energy distribution.
The question is when to stop the iteration for the
weight factor determination.
Our criterion for a satisfactory weight factor is that 
as long as we do get a random walk in potential energy space,
the probability distribution $P_{\rm mu}(E)$ does not have to
be completely
flat with a tolerance of, say, an order of magnitude deviation.
In such a case, we usually 
perform with this weight 
factor a multicanonical simulation with high statistics
(production run) in order to get even better estimate
of the density of states. 
Let $N_{\rm mu}(E)$ be the histogram
of potential energy distribution 
$P_{\rm mu} (E)$ 
obtained by this production run.
The best estimate of the density of states can then be
given by the single-histogram reweighting
techniques \cite{FS1} as follows (see the proportionality
relation in Eq. (\ref{eqn5})):
\begin{equation} 
n(E)= \displaystyle{\frac{N_{\rm mu}(E)}{W_{\rm mu}(E)}}~.
\label{eqn10c}
\end{equation}
By substituting this quantity into Eq. (\ref{eqn10a}),
one can calculate ensemble averages of physical
quantity $A(E)$ as a function of temperature.
Moreover, ensemble averages of any physical quantity $A$
(including those that cannot be expressed as functions
of potential energy) at any temperature 
$T$ ($= 1/k_{\rm B} \beta$) can now be obtained as long as
one stores the ``trajectory'' of configurations (and $A$)
%$x(k)$ $(k=1, \cdots, n_0)$ 
from the production run.  
%Here, $n_0$ is the total number of configurations stored.
Namely, we have
\begin{equation} 
%<A>_T  =  \frac{ \displaystyle{ 
%\sum_{x_k} A(x_k) W_{\rm mu}^{-1} (E(x_k))
%\exp(-\beta E(x_k)) } }
%{\displaystyle{ 
%\sum_{x_k} W_{\rm mu}^{-1} (E(x_k)) \exp(-\beta E(x_k)) } }~, 
<A>_T  =  \frac{ \displaystyle{ 
\sum_{k=1}^{n_0} A(x(k)) W_{\rm mu}^{-1} (E(x(k)))
\exp \left[-\beta E(x(k)) \right] } }
{\displaystyle{ 
\sum_{k=1}^{n_0} W_{\rm mu}^{-1} (E(x(k))) 
\exp \left[-\beta E(x(k)) \right] } }~, 
\label{eqn10d}
\end{equation}
where $x(k)$ is the configuration at the
$k$-th MC (or MD) step and $n_0$ is the total number 
of configurations stored. 
Note that when $A$ is a function of $E$, Eq.~(\ref{eqn10d}) reduces to
Eq.~(\ref{eqn10a}) where the density of states is given by
Eq.~(\ref{eqn10c}).
%are equivalent if we define
%\begin{equation}
%A(E) = \frac{1}{N_{\rm mu}(E)} \sum_{k^{\prime}} A_{k^{\prime}}~,
%\label{eqn10e}
%\end{equation}
%where the summation is taken over trajectory of the configurations
%with $E_{k^{\prime}} = E$ (this means that $E_{k^{\prime}}$
%belongs to the same bin as that defined by $E$; namely,
%$|E - E_{k^{\prime}}| < \epsilon$.

Eqs. (\ref{eqn10a}) and (\ref{eqn10d}) or any other equations
which involve summations of
exponential functions often encounter with numerical difficulties
such as overflows.  These can be overcome by using, 
for instance,
the following equation \cite{BergTXT,BergLog}:
For $C=A+B$ (with $A>0$ and $B>0$) we have
\begin{equation}
\begin{array}{rl} 
\ln C &= \ln \left[{\rm max}(A,B) \left(1 +
 \displaystyle{\frac{{\rm min}(A,B)}{{\rm max}(A,B)}} 
 \right) \right] \\
 &= {\rm max}(\ln A, \ln B) +
 \ln \left\{1+\exp \left[{\rm min}(\ln A,\ln B) -
 {\rm max}(\ln A,\ln B) \right] \right\}~.
\end{array}
\label{eqn10e}
\end{equation}

We now briefly review the original {\it simulated tempering} (ST)
method \cite{ST1,ST2}.  In this method temperature itself becomes a
dynamical variable, and both the configuration and the temperature are updated
during the simulation with a weight:
\begin{equation}
%W_{\rm ST} (E;T) = e^{-\beta E + a(T)}~,
W_{\rm ST} (E;T) = \exp \left(-\beta E + a(T) \right)~,
\label{Eqn1}
\end{equation}
where the function $a(T)$ is chosen so that the probability distribution
of temperature is flat:
\begin{equation}
P_{\rm ST}(T) = \int dE~ n(E)~ W_{{\rm ST}} (E;T) =
%\int dE~ n(E)~ e^{-\beta E + a(T)} = {\rm constant}~.
\int dE~ n(E)~ \exp \left(-\beta E + a(T) \right) = {\rm constant}~.
\lab{Eqn2}
\end{equation}
Hence, in simulated tempering the {\it temperature} is sampled
uniformly. A free random walk in temperature space
is realized, which in turn
induces a random walk in potential energy space and
allows the simulation to escape from
states of energy local minima.

In the numerical work we discretize the temperature in
$M$ different values, $T_m$ ($m=1, \cdots, M$).  Without loss of
generality we can order the temperature
so that $T_1 < T_2 < \cdots < T_M$.  The lowest temperature
$T_1$ should be sufficiently low so that the simulation can explore the
global-minimum-energy region, and
the highest temperature $T_M$ should be sufficiently high so that
no trapping in an energy-local-minimum state occurs.  The probability
weight factor in Eq.~(\ref{Eqn1}) is now written as
\begin{equation}
%W_{\rm ST}(E;T_m) = e^{-\beta_m E + a_m}~,
W_{\rm ST}(E;T_m) = \exp (-\beta_m E + a_m)~,
\label{Eqn3}
\end{equation}
where $a_m=a(T_m)$ ($m=1, \cdots, M$).
%The parameters $a_m$ are not known {\it a priori} and 
%have to be determined
%by iterations of short simulations.
%This process can be non-trivial and very difficult for complex systems.
Note that from Eqs.~(\ref{Eqn2}) and (\ref{Eqn3}) we have
\begin{equation}
%e^{-a_m} \propto \int dE~ n(E)~ e^{- \beta_m E}~.
\exp (-a_m) \propto \int dE~ n(E)~ \exp (- \beta_m E)~.
\label{Eqn4}
\end{equation}
The parameters $a_m$ are therefore ``dimensionless'' Helmholtz free energy
at temperature $T_m$
(i.e., the inverse temperature $\beta_m$ multiplied by
the Helmholtz free energy).
We remark that the density of states $n(E)$ (and hence, the
multicanonical weight factor) and the simulated tempering
weight factor $a_m$ are related by a Laplace transform
\cite{HO96a}.
The knowledge of one implies that of the other, although
in numerical work the inverse Laplace transform of 
Eq. (\ref{Eqn4}) is nontrivial.

Once the parameters $a_m$ are determined and the initial configuration and
the initial temperature $T_m$ are chosen,
a simulated tempering simulation is then realized by alternately
performing the following two steps \cite{ST1,ST2}:
\begin{enumerate}
\item A canonical MC or MD simulation at the fixed temperature $T_m$
(based on Eq. (\ref{eqn4d}) or Eq. (\ref{eqn4e})) is carried out 
for a certain steps.
\item The temperature $T_m$ is updated to the neighboring values
$T_{m \pm 1}$ with the configuration fixed.  The transition probability of
this temperature-updating
process is given by the Metropolis criterion (see Eq.~(\ref{Eqn3})):
\begin{equation}
w(T_m \rightarrow T_{m \pm 1})
= {\rm min}\left(1,\exp \left( - \Delta \right)\right)~,
%= \left\{
%\begin{array}{ll}
% 1~, & {\rm for} \ \Delta \le 0~, \cr
% \exp \left( - \Delta \right)~, & {\rm for} \ \Delta > 0~,
%\end{array}
%\right.
\label{Eqn5}
\end{equation}
where
\begin{equation}
\Delta = \left(\beta_{m \pm 1} - \beta_m \right) E
- \left(a_{m \pm 1} - a_m \right)~.
\label{Eqn6}
\end{equation}
\end{enumerate}
Note that in Step 2 we exchange only pairs of 
neighboring temperatures in order to secure sufficiently
large acceptance ratio of temperature updates.

As in multicanonical algorithm, the simulated tempering
parameters $a_m=a(T_m)$ ($m=1, \cdots, M$)
are also determined by iterations of short trial simulations
(see, e.g.,  Refs.~\cite{STrev,IRB1,HO96b} for details).
This process can be non-trivial and very tedius for complex
systems.

After the optimal simulated tempering weight factor is determined,
one performs a long simulated tempering run once.
The canonical expectation value of a physical quantity $A$
at temperature $T_m$ ($m=1, \cdots, M$) can be calculated
by the usual arithmetic mean as follows:
\begin{equation}
<A>_{T_m} = \frac{1}{n_{m}} \sum_{k=1}^{n_{m}}
A\left(x_{m}(k)\right)~,
\label{Eqn7}
\end{equation}
where $x_m(k)$ ($k=1,\cdots,n_m$) are the configurations 
obtained at temperature $T_m$
and $n_{m}$ is the total number of measurements made
at $T=T_m$.
The expectation value at any intermediate temperature
can also be obtained from
Eq.~(\ref{eqn10a}), where the 
density of states is given by
the multiple-histogram reweighting techniques \cite{FS2,WHAM}
as follows.
Let $N_m(E)$ and $n_m$ be respectively
the potential-energy histogram and the total number of
samples obtained at temperature $T_m=1/k_{\rm B} \beta_m$
($m=1, \cdots, M$). 
The best estimate of the density of states is then given by \cite{FS2,WHAM}
\begin{equation}
n(E) = \frac{\dis{\sum_{m=1}^M ~g_m^{-1}~N_m(E)}}
%{\dis{\sum_{m=1}^M ~g_m^{-1}~n_m~e^{f_m-\beta_m E}}}~,
{\dis{\sum_{m=1}^M ~g_m^{-1}~n_m~\exp (f_m-\beta_m E)}}~,
\label{Eqn8a}
\end{equation}
where we have for each $m$ ($=1, \cdots, M$)
\begin{equation}
%e^{-f_m} = \sum_{E} ~n(E)~e^{-\beta_m E}~.
\exp (-f_m) = \sum_{E} ~n(E)~\exp (-\beta_m E)~.
\label{Eqn8b}
\end{equation}
Here, $g_m = 1 + 2 \tau_m$,
and $\tau_m$ is the integrated
autocorrelation time at temperature $T_m$.
For many systems the quantity $g_m $ can safely 
be set to be a constant in the reweighting 
formulae \cite{WHAM}, and so we usually set $g_m =1$. 

Note that
Eqs.~(\ref{Eqn8a}) and
(\ref{Eqn8b}) are solved self-consistently
by iteration \cite{FS2,WHAM} to obtain
the density of states $n(E)$ and
the dimensionless Helmholtz free energy $f_m$.
Namely, we
can set all the $f_m$ ($m=1, \cdots, M$) to, e.g., zero initially.
We then use Eq.~(\ref{Eqn8a}) to obtain 
$n(E)$, which is substituted into
Eq.~(\ref{Eqn8b}) to obtain next values of $f_m$, and so on.

Moreover, ensemble averages of any physical quantity $A$
(including those that cannot be expressed as functions
of potential energy) at any temperature 
$T$ ($= 1/k_{\rm B} \beta$) can now be obtained from
the ``trajectory'' of configurations of 
the production 
run.  Namely, we first obtain $f_m$ ($m=1, \cdots, M$) by solving
Eqs. (\ref{Eqn8a}) and (\ref{Eqn8b}) self-consistently, 
and then we have \cite{MSO03}
\begin{equation}
%<A>_T  =  \frac{ \displaystyle{ 
%\sum_{m=1}^{M} \sum_{x_m} A(x_m)  \frac{ g_m^{-1} }
%{\displaystyle{\sum_{\ell=1}^{M} g_{\ell}^{-1} n_{\ell} 
%\exp(f_{\ell} - \beta_{\ell} E(x_m))}}
%\exp(-\beta E(x_m)) }}
%{\displaystyle{ 
%\sum_{m=1}^{M} \sum_{x_m} \frac{ g_m^{-1}  }
%{\displaystyle{\sum_{\ell=1}^{M} g_{\ell}^{-1} n_{\ell} 
%\exp(f_{\ell} - \beta_{\ell} E(x_m))}}
%\exp(-\beta E(x_m)) }}~,
%\label{eqn17}
<A>_T  =  \frac{ \displaystyle{ 
\sum_{m=1}^{M} \sum_{k=1}^{n_m} A(x_m(k))  \frac{ g_m^{-1} }
{\displaystyle{\sum_{\ell=1}^{M} g_{\ell}^{-1} n_{\ell} 
\exp \left[ f_{\ell} - \beta_{\ell} E(x_m(k)) \right]}}
\exp \left[ -\beta E(x_m(k)) \right] }}
{\displaystyle{ 
\sum_{m=1}^{M} \sum_{k=1}^{n_m} \frac{ g_m^{-1}  }
{\displaystyle{\sum_{\ell=1}^{M} g_{\ell}^{-1} n_{\ell} 
\exp \left[ f_{\ell} - \beta_{\ell} E(x_m(k)) \right]}}
\exp \left[ -\beta E(x_m(k)) \right] }}~,
\label{eqn17}
\end{equation}
where $x_m(k)$ ($k=1,\cdots,n_m$) are the configurations 
obtained at temperature $T_m$. \\ 
%Here, the trajectories $x_m$ are 
%taken for each temperature $T_m$ separately. 

\subsection{Replica-Exchange Method}

The {\it replica-exchange method} (REM) \cite{RE1}--\cite{RE2}
was developed as an extension of
simulated tempering \cite{RE1} (thus it is also referred to as
{\it parallel tempering} \cite{STrev})
(see, e.g.,  Ref.~\cite{SO} for a detailed
description of the algorithm).
The system for REM consists of 
$M$ {\it non-interacting} copies (or, replicas) 
of the original system in the canonical ensemble
at $M$ different temperatures $T_m$ ($m=1, \cdots, M$).
We arrange the replicas so that there is always
exactly one replica at each temperature.
Then there exists a one-to-one correspondence between replicas
and temperatures; the label $i$ ($i=1, \cdots, M$) for replicas 
is a permutation of 
the label $m$ ($m=1, \cdots, M$) for temperatures,
and vice versa:
\begin{equation}
\left\{
\begin{array}{rl}
i &=~ i(m) ~\equiv~ f(m)~, \cr
m &=~ m(i) ~\equiv~ f^{-1}(i)~,
\end{array}
\right.
\label{eq4b}
\end{equation}
where $f(m)$ is a permutation function of $m$ and
$f^{-1}(i)$ is its inverse.

Let $X = \left\{x_1^{[i(1)]}, \cdots, x_M^{[i(M)]}\right\} 
= \left\{x_{m(1)}^{[1]}, \cdots, x_{m(M)}^{[M]}\right\}$ 
stand for a ``state'' in this generalized ensemble.
%The state $X$ is specified by the $M$ sets of 
Each ``substate'' $x_m^{[i]}$ is specified by the
coordinates $q^{[i]}$ and momenta $p^{[i]}$
of $N$ atoms in replica $i$ at temperature $T_m$:
\begin{equation}
x_m^{[i]} \equiv \left(q^{[i]},p^{[i]}\right)_m~.
\label{eq5}
\end{equation}

Because the replicas are non-interacting, the weight factor for
the state $X$ in
this generalized ensemble is given by
the product of Boltzmann factors for each replica (or at each
temperature):
\begin{equation}
W_{\rm REM}(X) = \exp \left\{- \dis{\sum_{i=1}^M \beta_{m(i)} 
H\left(q^{[i]},p^{[i]}\right) } \right\}
 = \exp \left\{- \dis{\sum_{m=1}^M \beta_m 
H\left(q^{[i(m)]},p^{[i(m)]}\right) }
 \right\}~,
\label{eq7}
\end{equation}
where $i(m)$ and $m(i)$ are the permutation functions in 
Eq.~(\ref{eq4b}).

We now consider exchanging a pair of replicas in the generalized
ensemble.  Suppose we exchange replicas $i$ and $j$ which are
at temperatures $T_m$ and $T_n$, respectively:  
\begin{equation}
X = \left\{\cdots, x_m^{[i]}, \cdots, x_n^{[j]}, \cdots \right\} 
\longrightarrow \ 
X^{\prime} = \left\{\cdots, x_m^{[j] \prime}, \cdots, x_n^{[i] \prime}, 
\cdots \right\}~. 
\label{eq8}
\end{equation}
Here, $i$, $j$, $m$, and $n$ are related by the permutation
functions in Eq.~(\ref{eq4b}),
and the exchange of replicas introduces a new 
permutation function $f^{\prime}$:
\begin{equation}
\left\{
\begin{array}{rl}
i &= f(m) \longrightarrow j=f^{\prime}(m)~, \cr
j &= f(n) \longrightarrow i=f^{\prime}(n)~. \cr
\end{array}
\right.
\label{eq8c}
\end{equation}

The exchange of replicas can be written in more detail as
\begin{equation}
\left\{
\begin{array}{rl}
x_m^{[i]} \equiv \left(q^{[i]},p^{[i]}\right)_m & \longrightarrow \ 
x_m^{[j] \prime} \equiv \left(q^{[j]},p^{[j] \prime}\right)_m~, \cr
x_n^{[j]} \equiv \left(q^{[j]},p^{[j]}\right)_n & \longrightarrow \ 
x_n^{[i] \prime} \equiv \left(q^{[i]},p^{[i] \prime}\right)_n~,
\end{array}
\right.
\label{eq9}
\end{equation}
where the definitions for $p^{[i] \prime}$ and $p^{[j] \prime}$
will be given below.
We remark that this process is equivalent to exchanging
a pair of temperatures $T_m$ and $T_n$ for the
corresponding replicas $i$ and $j$ as follows:
\begin{equation}
\left\{
\begin{array}{rl}
x_m^{[i]} \equiv \left(q^{[i]},p^{[i]}\right)_m & \longrightarrow \ 
x_n^{[i] \prime} \equiv \left(q^{[i]},p^{[i] \prime}\right)_n~, \cr
x_n^{[j]} \equiv \left(q^{[j]},p^{[j]}\right)_n & \longrightarrow \ 
x_m^{[j] \prime} \equiv \left(q^{[j]},p^{[j] \prime}\right)_m~.
\end{array}
\right.
\label{eq10}
\end{equation}

In the original implementation of the 
{\it replica-exchange method} (REM) \cite{RE1}--\cite{RE2},
Monte Carlo algorithm was used, and only the coordinates $q$
(and the potential energy
function $E(q)$)
had to be taken into account.  
In molecular dynamics algorithm, on the other hand, we also have to
deal with the momenta $p$.
We proposed the following momentum 
assignment in Eq.~(\ref{eq9}) (and in Eq.~(\ref{eq10})) \cite{SO}:
\begin{equation}
\left\{
\begin{array}{rl}
p^{[i] \prime} & \equiv \dis{\sqrt{\frac{T_n}{T_m}}} ~p^{[i]}~, \cr
p^{[j] \prime} & \equiv \dis{\sqrt{\frac{T_m}{T_n}}} ~p^{[j]}~,
\end{array}
\right.
\label{eq11}
\end{equation}
which we believe is the simplest and the most natural.
This assignment means that we just rescale uniformly 
the velocities of all the atoms 
in the replicas by
the square root of the ratio of the two temperatures so that
the temperature condition in Eq.~(\ref{eqn4}) may be satisfied.

In order for this exchange process to converge towards an equilibrium
distribution, it is sufficient to impose the detailed balance
condition on the transition probability $w(X \rightarrow X^{\prime})$:
\begin{equation}
\frac{W_{\rm REM}(X)}{Z} \  w(X \rightarrow X^{\prime})
= \frac{W_{\rm REM}(X^{\prime})}{Z} \  w(X^{\prime} \rightarrow X)~,
\label{eq12}
\end{equation}
where $Z$ is the partition function of the entire system.
From Eqs.~(\ref{eqn1}), (\ref{eqn2}), (\ref{eq7}), (\ref{eq11}), 
and (\ref{eq12}), we have
\begin{equation}
\begin{array}{rl}
\dis{\frac{w(X \rightarrow X^{\prime})} 
     {w(X^{\prime} \rightarrow X)}} 
&= \exp \left\{ 
- \beta_m \left[K\left(p^{[j] \prime}\right) + E\left(q^{[j]}\right)\right] 
- \beta_n \left[K\left(p^{[i] \prime}\right) + E\left(q^{[i]}\right)\right]
\right. \cr
& \ \ \ \ \ \ \ \ \ \  \left.
+ \beta_m \left[K\left(p^{[i]}\right) + E\left(q^{[i]}\right)\right] 
+ \beta_n \left[K\left(p^{[j]}\right) + E\left(q^{[j]}\right)\right]
\right\}~, \cr
&= \exp \left\{ 
- \beta_m \dis{\frac{T_m}{T_n}} K\left(p^{[j]}\right)
- \beta_n \dis{\frac{T_n}{T_m}} K\left(p^{[i]}\right)
+ \beta_m K\left(p^{[i]}\right)
+ \beta_n K\left(p^{[j]}\right)
\right. \cr
& \ \ \ \ \ \ \ \ \ \  \left.
- \beta_m \left[E\left(q^{[j]}\right)
                - E\left(q^{[i]}\right)\right] 
- \beta_n \left[E\left(q^{[i]}\right)
                - E\left(q^{[j]}\right)\right] 
\right\}~, \cr
&= \exp \left( - \Delta \right)~,
\end{array}
\label{eq13}
\end{equation}
where
%\begin{equation}
%\Delta \equiv \left(\beta_n - \beta_m \right)
%              \left(E\left(q^{[i]}\right)
%                  - E\left(q^{[j]}\right)\right)~, 
%\label{eq14}
%\end{equation}
\begin{eqnarray}
\Delta &=& \beta_m 
\left(E\left(q^{[j]}\right) - E\left(q^{[i]}\right)\right) 
- \beta_n
\left(E\left(q^{[j]}\right) - E\left(q^{[i]}\right)\right)~,
\label{eqn14a} \\
  &=& \left(\beta_m - \beta_n \right)
\left(E\left(q^{[j]}\right) - E\left(q^{[i]}\right)\right)~, 
\label{eqn14b}
\end{eqnarray}
and $i$, $j$, $m$, and $n$ are related by the permutation
functions in Eq.~(\ref{eq4b}) before the exchange:
\begin{equation}
\left\{
\begin{array}{ll}
i &= f(m)~, \cr
j &= f(n)~.
\end{array}
\right.
\label{eq13b}
\end{equation}
This can be satisfied, for instance, by the usual Metropolis criterion
\cite{Metro}:
\begin{equation}
w(X \rightarrow X^{\prime}) \equiv
w\left( x_m^{[i]} ~\left|~ x_n^{[j]} \right. \right) 
= {\rm min}\left(1,\exp \left( - \Delta \right)\right)~,
%= \left\{
%\begin{array}{ll}
% 1~, & {\rm for} \ \Delta \le 0~, \cr
% \exp \left( - \Delta \right)~, & {\rm for} \ \Delta > 0~,
%\end{array}
%\right.
\label{eq15}
\end{equation}
where in the second expression 
(i.e., $w( x_m^{[i]} | x_n^{[j]} )$) 
we explicitly wrote the
pair of replicas (and temperatures) to be exchanged.
Note that this is exactly the same criterion that was originally
derived for Monte Carlo algorithm \cite{RE1}--\cite{RE2}.

Without loss of generality we can
again assume $T_1 < T_2 < \cdots < T_M$.
A simulation of the 
{\it replica-exchange method} (REM) \cite{RE1}--\cite{RE2}
is then realized by alternately performing the following two
steps:
\begin{enumerate}
\item Each replica in canonical ensemble of the fixed temperature 
is simulated $simultaneously$ and $independently$
for a certain MC or MD steps. 
\item A pair of replicas at neighboring temperatures,
say $x_m^{[i]}$ and $x_{m+1}^{[j]}$, are exchanged
with the probability
$w\left( x_m^{[i]} ~\left|~ x_{m+1}^{[j]} \right. \right)$ 
in Eq.~(\ref{eq15}).
\end{enumerate}
Note that in Step 2 we exchange only pairs of replicas corresponding to
neighboring temperatures, because
the acceptance ratio of the exchange process decreases exponentially
with the difference of the two $\beta$'s (see Eqs.~(\ref{eqn14b})
and (\ref{eq15})).
Note also that whenever a replica exchange is accepted
in Step 2, the permutation functions in Eq.~(\ref{eq4b})
are updated.

The REM simulation is particularly suitable for parallel
computers.  Because one can minimize the amount of information
exchanged among nodes, it is best to assign each replica to
each node (exchanging pairs of temperature values among nodes
is much faster than exchanging coordinates and momenta).
This means that we keep track of the permutation function
$m(i;t)=f^{-1}(i;t)$ in Eq.~(\ref{eq4b}) as a function
of MC or MD step $t$ during the simulation.
After parallel canonical MC or MD simulations for a certain
steps (Step 1), $M/2$ pairs of
replicas corresponding to neighboring temperatures
are simulateneously exchanged (Step 2), and the pairing is alternated 
between the two possible choices, i.e., ($T_1,T_2$), ($T_3,T_4$), $\cdots$
and ($T_2,T_3$), ($T_4,T_5$), $\cdots$.

The major advantage of REM over other generalized-ensemble
methods such as multicanonical algorithm \cite{MUCA1,MUCA2}
and simulated tempering \cite{ST1,ST2}
lies in the fact that the weight factor 
is {\it a priori} known (see Eq.~(\ref{eq7})), while
in the latter algorithms the determination of the
weight factors can be very tedius and time-consuming.
A random walk in ``temperature space'' is
realized for each replica, which in turn induces a random
walk in potential energy space.  This alleviates the problem
of getting trapped in states of energy local minima.
%For the optimal performance of REM, however, one still has
%to choose an appropriate temperature distribution.
%There exists an iterative procedure for this \cite{RE1},
%and we have modified it further.  The details of our procedure 
%for the determination of the optimal temperature distribution
%will be given elsewhere \cite{SO}.
In REM, however, the number of required replicas increases
as the system size $N$ increases (according to $\sqrt N$) \cite{RE1}.
This demands a lot of computer power for complex systems. 

\subsection{Replica-Exchange Multicanonical Algorithm and
Replica-Exchange Simulated Tempering} 

The {\it replica-exchange multicanonical algorithm} (REMUCA) 
\cite{SO3,MSO03} overcomes
both the difficulties of MUCA (the multicanonical weight factor
determination is non-trivial)
and REM (a lot of replicas, or computation time, is required).
In REMUCA we first perform a short REM simulation (with $M$ replicas)
to determine the
multicanonical weight factor and then perform with this weight
factor a regular multicanonical simulation with high statistics.
The first step is accomplished by the multiple-histogram reweighting
techniques \cite{FS2,WHAM}.
Let $N_m(E)$ and $n_m$ be respectively
the potential-energy histogram and the total number of
samples obtained at temperature $T_m$ ($=1/k_{\rm B} \beta_m$) 
of the REM run.
The density of states $n(E)$ is then given by solving 
Eqs.~(\ref{Eqn8a}) and (\ref{Eqn8b}) self-consistently by iteration.
%\cite{FS2,WHAM}.

Once the estimate of the density of states is obtained, the
multicanonical weight factor can be directly determined from
Eq.~(\ref{eqn6}) (see also Eq.~(\ref{eqn7})).
Actually, the density of states $n(E)$ and 
the multicanonical potential energy, $E_{\rm mu}(E;T_0)$,
thus determined are only reliable in the following range:
\begin{equation}
E_1 \le E \le E_M~,
\label{eqn29}
\end{equation}
where 
\begin{equation}
\left\{
\begin{array}{rl}
E_1 &=~ <E>_{T_1}~, \\
E_M &=~ <E>_{T_M}~,
\end{array}
\right.
\label{eqn29b}
\end{equation}
and $T_1$ and $T_M$ are respectively the lowest and the highest
temperatures used in the REM run.
Outside this range we extrapolate
the multicanonical potential energy linearly: \cite{SO3}
\begin{equation}
 {\cal E}_{\rm mu}^{\{0\}}(E) \equiv \left\{
   \begin{array}{@{\,}ll}
   \left. \dis{\frac{\partial E_{\rm mu}(E;T_0)}{\partial E}}
        \right|_{E=E_1} (E - E_1)
             + E_{\rm mu}(E_1;T_0)~, &
         \mbox{for $E < E_1$,} \\
         E_{\rm mu}(E;T_0)~, &
         \mbox{for $E_1 \le E \le E_M$,} \\
   \left. \dis{\frac{\partial E_{\rm mu}(E;T_0)}{\partial E}}
        \right|_{E=E_M} (E - E_M)
             + E_{\rm mu}(E_M;T_0)~, &
         \mbox{for $E > E_M$.}
   \end{array}
   \right.
\label{eqn31}
\end{equation}
%In the present work the multicanonical potential
%energy function, $E_{\rm mu}(E;T_0)$, and its derivative,
%$\frac{\partial E_{\rm mu}(E;T_0)}{\partial E}$, were obtained by
%fitting the density of states, $n(E)$, (which was
%determined by the multiple-histogram reweighting
%techniques) by the cubic
%spline functions in the energy range of Eq.~(\ref{eqn29}).
The multicanonical MC and MD runs  
are then performed respectively with
the Metropolis criterion of Eq.~(\ref{eqn8})
and with the modified Newton equation 
in Eq.~(\ref{eqn9a}), 
in which 
${\cal E}_{\rm mu}^{\{0\}}(E)$ in
Eq.~(\ref{eqn31}) is substituted into $E_{\rm mu}(E;T_0)$.
We expect to obtain a flat potential energy distribution in
the range of Eq.~(\ref{eqn29}).
Finally, the results are analyzed by the single-histogram
reweighting techniques as described in Eq.~(\ref{eqn10c})
(and Eq.~(\ref{eqn10a})).

Some remarks are now in order.
From Eqs.~(\ref{eqn7}), (\ref{eqn9c}), (\ref{eqn9d}),
and (\ref{eqn29b}), Eq.~(\ref{eqn31}) becomes
\begin{equation}
 {\cal E}_{\rm mu}^{\{0\}}(E) = \left\{
   \begin{array}{@{\,}ll}
    \dis{\frac{T_0}{T_1}} (E - E_1) + T_0 S(E_1) = 
    \dis{\frac{T_0}{T_1}} E + {\rm constant}~, &
         \mbox{for $E < E_1 \equiv <E>_{T_1}$,} \\
         T_0 S(E)~, &
         \mbox{for $E_1 \le E \le E_M$,} \\
    \dis{\frac{T_0}{T_M}} (E - E_M) + T_0 S(E_M) = 
    \dis{\frac{T_0}{T_M}} E + {\rm constant}~, &
         \mbox{for $E > E_M \equiv <E>_{T_M}$.}
   \end{array}
   \right.
\label{eqn31b}
\end{equation}
The Newton equation in Eq.~(\ref{eqn9a}) is then written as
(see Eqs.~(\ref{eqn9b}), (\ref{eqn9c}), and (\ref{eqn9d}))
\begin{equation}
\dot{{\bip}}_k = \left\{
   \begin{array}{@{\,}ll}
   \dis{\frac{T_0}{T_1}}~{\bif}_k~, &
         \mbox{for $E < E_1$,} \\
   \dis{\frac{T_0}{T(E)}}~{\bif}_k~, &
         \mbox{for $E_1 \le E \le E_M$,} \\
   \dis{\frac{T_0}{T_M}}~{\bif}_k~, &
         \mbox{for $E > E_M$.}
   \end{array}
   \right.
\label{eqn31c}
\end{equation}
Because only the product of inverse temperature $\beta$ and
potential energy $E$ enters in the Boltzmann factor
(see Eq.~(\ref{eqn4b})), a rescaling of the potential energy
(or force) by a constant, say $\alpha$, can be considered as
%the rescaling of the temperature by $\alpha^{-1}$ \cite{HOE96,Yama}.
the rescaling of the temperature by $1/\alpha$ \cite{HOE96,Yama}.  
Hence,
our choice of ${\cal E}_{\rm mu}^{\{0\}}(E)$
in Eq.~(\ref{eqn31}) results in a canonical simulation at
$T=T_1$ for $E < E_1$, a multicanonical simulation for
$E_1 \le E \le E_M$, and a canonical simulation at
$T=T_M$ for $E > E_M$.
Note also that the above arguments are independent of
the value of $T_0$, and we
will get the same results, regardless of its value.

For Monte Carlo method, the above statement follows
directly from the following equation.  Namely, our
choice of the multicanonical potential energy in
Eq.~(\ref{eqn31}) gives (by substituting
Eq.~(\ref{eqn31b}) into Eq.~(\ref{eqn6}))
\begin{equation}
W_{\rm mu}(E) = \exp \left[-\beta_0 {\cal E}_{\rm mu}^{\{0\}}(E) \right]
 = \left\{
   \begin{array}{@{\,}ll}
   \dis{\exp \left(-\beta_1 E + {\rm constant}\right)}~, &
         \mbox{for $E < E_1$,} \\
   \dis{\frac{1}{n(E)}}~, &
         \mbox{for $E_1 \le E \le E_M$,} \\
   \dis{\exp \left(-\beta_M E + {\rm constant}\right)}~, &
         \mbox{for $E > E_M$.}
   \end{array}
   \right.
\label{eqn31d}
\end{equation}

%Since the WHAM equations are based on histograms,
%the density of states $n(E)$, or the multicanonical 
%potential energy 
%$E_{\rm mu}(E;T_0)$, will be given 
%in discrete
%values of the potential energy $E$.
%For multicanonical MD simulations, however, we need the
%derivative of $E_{\rm mu}(E;T_0)$ with respect to $E$
%(see Eq.~(\ref{eqn9a})).
%We thus introduce some smooth function to
%fit the data.  It is best to fit the derivative
%$\frac{\partial E_{\rm mu}(E;T_0)}{\partial E}$
%directly rather than $E_{\rm mu}(E;T_0)$ itself.
We now present another effective method of the multicanonical
weight factor \cite{RevSO}, which is closely related to
REMUCA.
We first perform a short REM simulation as in REMUCA
and calculate $<E>_{T}$ as a function of $T$
by the multiple-histogram
reweighting techniques (see Eqs.~(\ref{Eqn8a}) and (\ref{Eqn8b})).
Let us recall the Newton equation of
Eq.~(\ref{eqn9b}) and the thermodynamic
relation of Eqs.~(\ref{eqn9c}) and (\ref{eqn9d}).
The effective temperature $T(E)$, or the derivative
$\frac{\partial E_{\rm mu}(E;T_0)}{\partial E}$,
can be numerically obtained as the inverse function of
Eq.~(\ref{eqn9d}), 
where the average $<E>_{T(E)}$ has 
been obtained from the results of the REM simulation
by the multiple-histogram reweighting techniques.
Given its derivative, the multicanonical potential
energy can then be obtained by numerical integration 
(see Eqs.~(\ref{eqn7}) and (\ref{eqn9c})): \cite{RevSO}
\begin{equation}
E_{\rm mu}(E;T_0) = 
T_0 \int_{E_1}^{E} \frac{\partial S(E)}{\partial E} dE
= T_0 \int_{E_1}^{E} \frac{dE}{T(E)}~.
\label{EQ10}
\end{equation}
We remark that the same equation was used to obtain
the multicanonical weight factor in Ref.~\cite{H97c},
where $<E>_{T}$ was estimated by simulated
annealing instead of REM.
Essentially the same formulation was also recently
used in Ref.~\cite{TMK03}
to obtain the multicanonical potential energy,
where $<E>_{T}$ was calculated by conventional
canonical simulations.

%Finally, although we did not find any difficulty in the case 
%of protein systems that we studied,
%a single REM run in general may not be able to
%give an accurate estimate of the
%density of states (like in the case of
%a strong first-order phase transition \cite{RE1}).  In such a
%case we can still greatly simplify the process of the
%multicanonical weight factor determination by
%combining the present method with the
%previous iterative methods \cite{MUCA3,OH,KPV,SmBr,H97c,MUCAW,BK}.

We finally present the new method which we refer to as the 
{\it replica-exchange simulated tempering} (REST) \cite{MO4}.  
In this method, just as in REMUCA,
we first perform a short REM simulation (with $M$ replicas)
to determine the simulated tempering
weight factor and then perform with this weight
factor a regular ST simulation with high statistics.
The first step is accomplished by 
the multiple-histogram reweighting
techniques \cite{FS2,WHAM}, which give
the dimensionless Helmholtz free energy $f_m$ (see Eqs.~(\ref{Eqn8a})
and (\ref{Eqn8b})).

Once the estimate of the dimensionless Helmholtz free energy $f_m$ are
obtained, the simulated tempering 
weight factor can be directly determined by using
Eq.~(\ref{Eqn3}) where we set $a_m = f_m$ (compare Eq.~(\ref{Eqn4})
with Eq.~(\ref{Eqn8b})).
A long simulated tempering run is then performed with this
weight factor.  
Let $N_m(E)$ and $n_m$ be respectively
the potential-energy histogram and the total number of
samples obtained at temperature $T_m$ ($=1/k_{\rm B} \beta_m$) from this
simulated tempering run.  The multiple-histogram
reweighting techniques of Eqs.~(\ref{Eqn8a}) and (\ref{Eqn8b}) can be used
again to obtain the best estimate of the density of states
$n(E)$.
The expectation value of a physical quantity $A$
at any temperature $T~(= 1/k_{\rm B} \beta)$ is then calculated from
Eq.~(\ref{eqn10a}).

The formulations of REMUCA and REST are simple and straightforward, but
the numerical improvement is great, because the weight factor
determination for MUCA and ST becomes very difficult
by the usual iterative processes for complex systems.

\subsection{Multicanonical Replica-Exchange Method and
Simulated Tempering Replica-Exchange Method}

In the previous subsection we presented REMUCA, 
which uses a short REM run for the determination 
of the multicanonical weight factor. 
Here, we present two modifications of REM and refer the new methods as 
multicanonical replica-exchange method (MUCAREM) \cite{SO3,MSO03}
and simulated tempering replica-exchange method (STREM)
\cite{STREM}. 
In MUCAREM  the production run is 
a REM simulation with a few replicas
not in the canonical ensemble but
in the multicanonical ensemble, i.e.,
different replicas perform MUCA simulations with
different energy ranges.  Likewise in STREM
the production run is a REM simulation with a few replicas
that performs ST simulations with different temperature
ranges.
While MUCA and ST simulations are usually based on local
updates, a replica-exchange process can be considered to be
a global update, and global updates enhance the sampling
further.

We first describe MUCAREM.
Let ${\cal M}$ be the number of replicas.  Here, each replica
is in one-to-one correspondence not with temperature but with
multicanonical weight factors of different energy range.
Note that because multicanonical simulations cover much wider energy
ranges than regular canonical simulations, the number of
required replicas for the production run of MUCAREM is
much less than that for the regular REM (${\cal M} \ll M$).
The weight factor for this generalized ensemble
is now given by (see Eq. (\ref{eq7}))
\begin{equation}
W_{\rm MUCAREM} (X) = \displaystyle{ \prod_{i=1}^{{\cal M}} 
W_{\rm mu}^{\{m(i)\}}} \left( E \left(x_{m(i)}^{[i]} \right) \right) 
= \displaystyle{ \prod_{m=1}^{{\cal M}} 
W_{\rm mu}^{\{m\}}} \left( E \left( x_m^{[i(m)]} \right) \right)~,
\end{equation}
where we prepare the multicanonical weight factor (and the
density of states) separately for $m$ regions (see Eq.~(\ref{eqn6})):
\begin{equation}
%W_{\rm mu}^{\{m\}} \left( E \left(x_m^{[i(m)]} \right) \right) = 
%\exp \left[- \beta_m {\cal E}_{\rm mu}^{\{m\}}
%\left( E \left(x_m^{[i(m)]} \right) \right) \right]
%= \frec{1}{n^{\{m\}}\left(E\left(x_m^{[i(m)]}\right)\right)}~.
W_{\rm mu}^{\{m\}} \left( E \left(x_m^{[i]} \right) \right) = 
\exp \left[- \beta_m {\cal E}_{\rm mu}^{\{m\}}
\left( E \left(x_m^{[i]} \right) \right) \right]
\equiv \frac{1}{n^{\{m\}}\left(E\left(x_m^{[i]}\right)\right)}~.
\label{eqn23}
\end{equation}
Here, we have introduced ${\cal M}$ arbitrary 
reference temperatures
$T_m = 1/k_{\rm B} \beta_m$ ($m = 1, \cdots, {\cal M}$), but
the final results will be independent of the values of $T_m$, 
as one can see from the second equality in Eq.~(\ref{eqn23})
(these arbitrary 
temperatures are necessary only for MD simulations).
   
Each multicanonical weight factor 
$W_{\rm mu}^{\{m\}}(E)$, or the density of states
$n^{\{m\}}(E)$, 
is defined as follows. 
For each $m$ ($m=1,\cdots, {\cal M}$), 
we assign a pair of temperatures ($T_{\rm L}^{\{m\}},T_{\rm H}^{\{m\}}$). 
Here, we assume that $T_{\rm L}^{\{m\}} < T_{\rm H}^{\{m\}}$ and arrange the temperatures so that 
the neighboring regions covered by the pairs have sufficient overlaps. 
Without loss of generality we can assume 
$T_{\rm L}^{\{1\}} < \cdots < T_{\rm L}^{\{{\cal M}\}}$ and
$T_{\rm H}^{\{1\}} < \cdots < T_{\rm H}^{\{{\cal M}\}}$. 
We define the following quantities:
\begin{equation}
\left\{
\begin{array}{rl}
E_{\rm L}^{\{m\}} &=~ <E>_{{T_{\rm L}}^{\{m\}}}~, \\
E_{\rm H}^{\{m\}} &=~ <E>_{{T_{\rm H}}^{\{m\}}}~,~~ \mbox{($m=1,\cdots,{\cal M}$)}~.
\end{array}
\right.
\end{equation}

Suppose that the multicanonical weight factor $W_{\rm mu}(E)$
(or equivalently, the multicanonical potential energy
$E_{\rm mu}(E;T_0)$ in Eq. (\ref{eqn7}))
has been obtained as in REMUCA or by any other methods
in the entire energy range of interest ($E_{\rm L}^{\{1\}} < E < 
E_{\rm H}^{\{{\cal M}\}}$). 
We then have for each $m$ ($m=1,\cdots, {\cal M}$)
the following multicanonical potential energies
(see Eq. (\ref{eqn31})): \cite{SO3}
\begin{equation}
 {\cal E}_{\rm mu}^{\{m\}}(E) = \left\{
   \begin{array}{@{\,}ll}
   \left. \dis{\frac{\partial E_{\rm mu}(E;T_m)}{\partial E}}
        \right|_{E=E_{\rm L}^{\{m\}}} (E - E_{\rm L}^{\{m\}})
             + E_{\rm mu}(E_{\rm L}^{\{m\}};T_m)~, &
         \mbox{for $E < E_{\rm L}^{\{m\}}$,} \\
         E_{\rm mu}(E;T_m)~, &
         \mbox{for $E_{\rm L}^{\{m\}} \le E \le E_{\rm H}^{\{m\}}$,} \\
   \left. \dis{\frac{\partial E_{\rm mu}(E;T_m)}{\partial E}}
        \right|_{E=E_{\rm H}^{\{m\}}} (E - E_{\rm H}^{\{m\}})
             + E_{\rm mu}(E_{\rm H}^{\{m\}};T_m)~, &
         \mbox{for $E > E_{\rm H}^{\{m\}}$.}
   \end{array}
   \right. 
\label{eqn33}
\end{equation}
 
Finally, a MUCAREM simulation 
is realized by alternately performing the following two steps.
\begin{enumerate}
\item Each replica of the fixed multicanonical ensemble is simulated 
$simultaneously$ and $independently$ for a certain MC or MD steps.
\item A pair of replicas, say $i$ and $j$, which are in neighboring 
multicanonical ensembles, say $m$-th and $(m+1)$-th, 
respectively, are exchanged:
$X = \left\{\cdots, x_m^{[i]}, \cdots, x_{m+1}^{[j]}, \cdots \right\}
\longrightarrow \
X^{\prime} = \left\{\cdots, x_m^{[j]}, \cdots, x_{m+1}^{[i]},
\cdots \right\}$.
The transition probability of this replica exchange is 
given by the Metropolis criterion:
\begin{equation}
w(X \rightarrow X^{\prime}) 
= {\rm min}\left(1,\exp \left( - \Delta \right)\right)~,
\label{eqn28}
\end{equation}
where we now have (see Eq.~(\ref{eqn14a})) \cite{SO3}
\begin{equation}
%\Delta = \beta_{m+1}
%\left\{{\cal E}_{\rm mu}^{\{m+1\}}\left(E\left(q^{[i]}\right)\right) -
%{\cal E}_{\rm mu}^{\{m+1\}}\left(E\left(q^{[j]}\right)\right)\right\}
%- \beta_{m}
%\left\{{\cal E}_{\rm mu}^{\{m\}}\left(E\left(q^{[i]}\right)\right) -
%{\cal E}_{\rm mu}^{\{m\}}\left(E\left(q^{[j]}\right)\right)\right\}~.
\Delta = \beta_{m}
\left\{{\cal E}_{\rm mu}^{\{m\}}\left(E\left(q^{[j]}\right)\right) -
{\cal E}_{\rm mu}^{\{m\}}\left(E\left(q^{[i]}\right)\right)\right\}
- \beta_{m+1}
\left\{{\cal E}_{\rm mu}^{\{m+1\}}\left(E\left(q^{[j]}\right)\right) -
{\cal E}_{\rm mu}^{\{m+1\}}\left(E\left(q^{[i]}\right)\right)\right\}~.
\label{Eqn21}
\end{equation}
Here $E\left(q^{[i]}\right)$ and $E\left(q^{[j]}\right)$ are the potential energy 
of the $i$-th replica and the $j$-th replica, respectively.
\end{enumerate}

Note that in Eq. (\ref{Eqn21}) we need to newly evaluate the 
multicanonical
potential energy, ${\cal E}_{\rm mu}^{\{m\}}(E(q^{[j]}))$ and
${\cal E}_{\rm mu}^{\{m+1\}}(E(q^{[i]}))$, because 
${\cal E}_{\rm mu}^{\{m\}}(E)$ and
${\cal E}_{\rm mu}^{\{n\}}(E)$ are, in
general, different functions for $m \ne n$. 
%We remark that the same additional evaluation of the
%potential energy is necessary for the multidimensional replica-exchange
%method \cite{SKO}.

In this algorithm, the $m$-th multicanonical ensemble actually 
results in a 
canonical simulation at  
$T= T_{\rm L}^{\{m\}}$  
for $E < E_{\rm L}^{\{m\}}$, a multicanonical simulation for 
$E_{\rm L}^{\{m\}} \le E \le E_{\rm H}^{\{m\}}$, and 
a canonical simulation at 
$T=T_{\rm H}^{\{m\}}$ 
for $E > E_{\rm H}^{\{m\}}$, 
while the replica-exchange process samples states of 
the whole energy range 
($E_{\rm L}^{\{1\}} \le E \le E_{\rm H}^{\{{\cal M}\}}$).

For obtaining the canonical distributions at any
intermediate temperature $T$,
the multiple-histogram reweighting techniques \cite{FS2,WHAM}
are again used.
Let $N_m(E)$ and $n_m$ be respectively
the potential-energy histogram and the total number of
samples obtained 
%at $T_m$ with the multicanonical
%potential energy ${\cal E}_{\rm mu}^{\{m\}}(E)$
with the multicanonical weight factor
$W_{\rm mu}^{\{m\}}(E)$
($m = 1, \cdots, {\cal M}$).
The expectation value
of a physical quantity $A$ 
at any temperature $T$ ($=1/k_{\rm B} \beta$)
is then obtained from Eq.~(\ref{eqn10a}),
where the best estimate of the density of states is obtained by
solving the WHAM equations,
which now read \cite{SO3}
\begin{equation}
n(E) = \frac{  \displaystyle{ \sum_{m=1}^{\cal M} g_m^{-1} N_m (E)} }
{ \displaystyle{ \sum_{m=1}^{\cal M} g_m^{-1} n_m \exp(f_m) W_{\rm mu}^{\{m\}}(E) } }
= \frac{\dis{\sum_{m=1}^{\cal M} ~g_m^{-1}~N_m(E)}}
{\dis{\sum_{m=1}^{\cal M} 
%~g_m^{-1}~n_m~e^{f_m-\beta_m {\cal E}_{\rm mu}^{\{m\}}(E)}}}~,
~g_m^{-1}~n_m~\exp \left(f_m-\beta_m {\cal E}_{\rm mu}^{\{m\}}(E)\right)}}~,
\label{eqn30}
\end{equation}
and for each $m$ ($=1, \cdots, {\cal M}$)
\begin{equation}
\exp(-f_m) = \sum_E n(E) W_{\rm mu}^{\{m\}}(E)
%= \sum_{E} ~n(E)~e^{-\beta_m {\cal E}_{\rm mu}^{\{m\}}(E)}~.
= \sum_{E} ~n(E)~\exp \left(-\beta_m {\cal E}_{\rm mu}^{\{m\}}(E)\right)~.
\label{Eqn31}
\end{equation}
%Here, parameters are given in subsection of REM. 
Note that $W_{\rm mu}^{\{m\}}(E)$ is used 
instead of the Boltzmann factor $\exp (- \beta_m E)$ in
Eqs. (\ref{Eqn8a}) and (\ref{Eqn8b}). 
%$n(E)$ and $f_m$ are solved self-consistently by iteration. 
%$g_m = 1 + 2 \tau_m$, and $\tau_m$ is the integrated autocorrelation 
%time of multicanonical simulation for  
%$E_{\rm L}^{\{m\}} \le E \le E_{\rm H}^{\{m\}}$.
%For biomolecular systems the quantity $g_m $ can safely be set to 
%be a constant in the reweighting 
%formulae, \cite{WHAM} and so we set $g_m =1 $ throughout the analyses in 
%the present work. 

Moreover, ensemble averages of any physical quantity $A$
(including those that cannot be expressed as functions
of potential energy) at any temperature 
$T$ ($= 1/k_{\rm B} \beta$) can now be obtained from
the ``trajectory'' of configurations of 
the production 
run.  Namely, we first obtain $f_m$ ($m=1, \cdots, {\cal M}$) by solving
Eqs. (\ref{eqn30}) and (\ref{Eqn31}) self-consistently, 
and then we have \cite{MSO03}
\begin{equation} 
%<A>_T  =  \frac{ \displaystyle{ 
%\sum_{m=1}^{\cal M} \sum_{x_m} A(x_m)  \frac{ g_m^{-1} }
%{\displaystyle{\sum_{\ell=1}^{\cal M} g_{\ell}^{-1} n_{\ell} 
%\exp(f_{\ell}) W_{\rm mu}^{\{m\}}(E(x_m))  }}
%\exp \left(-\beta E(x_m) \right) }}
%{\displaystyle{ 
%\sum_{m=1}^{\cal M} \sum_{x_m} \frac{ g_m^{-1}  }
%{\displaystyle{\sum_{\ell=1}^{\cal M} g_{\ell}^{-1} n_{\ell} 
%\exp(f_{\ell}) W_{\rm mu}^{\{m\}}(E(x_m))  }}
%\exp \left(-\beta E(x_m) \right) }}~,
<A>_T  =  \frac{ \displaystyle{ 
\sum_{m=1}^{\cal M} \sum_{k=1}^{n_m} A(x_m(k))  \frac{ g_m^{-1} }
{\displaystyle{\sum_{\ell=1}^{\cal M} g_{\ell}^{-1} n_{\ell} 
\exp(f_{\ell}) W_{\rm mu}^{\{m\}}(E(x_m(k)))  }}
\exp \left[-\beta E(x_m(k)) \right] }}
{\displaystyle{ 
\sum_{m=1}^{\cal M} \sum_{k=1}^{n_m} \frac{ g_m^{-1}  }
{\displaystyle{\sum_{\ell=1}^{\cal M} g_{\ell}^{-1} n_{\ell} 
\exp(f_{\ell}) W_{\rm mu}^{\{m\}}(E(x_m(k)))  }}
\exp \left[-\beta E(x_m(k)) \right] }}~,
\label{eqn32}
\end{equation}
where the trajectories $x_m(k)$ ($k=1,\cdots,n_m$) are 
taken from each multicanonical simulation 
with the multicanonical weight factor
$W_{\rm mu}^{\{m\}}(E)$ ($m=1, \cdots ,{\cal M}$) separately. \\
%for the energy range of 
%$E_{\rm L}^{\{m\}} \le E \le E_{\rm H}^{\{m\}}$
%separately.\\ 
%Note that the formation of Eqs. (\ref{eqn30}), (\ref{eqn31}), and (\ref{eqn32}) are 
%the general formula, if we perform the simulation with 
%the weight factor given by the product of 
%$W^{\{m\}} (E(x_m))$ ($m=1, \cdots ,{\cal M}$). \\

As seen above, both REMUCA and MUCAREM can be used to obtain the
multicanonical weight factor, or the density of states,
for the entire potential energy range of interest.
For complex systems, however, a single REMUCA or MUCAREM
simulation is often insufficient.
In such cases we can iterate MUCA (in REMUCA) and/or 
MUCAREM simulations
in which the estimate of the multicanonical weight factor
is updated
by the single- and/or multiple-histogram reweighting techniques,
respectively.
%The MUCA production run (in REMUCA) or MUCAREM
%production run with the refined multicanonical weight factor 
%will make a more frequent random walk in energy space and 
%can yield a flatter probability distribution of potential energy. 
%%to accurate thermodynamic quantities. 

%Sometimes we obtain the multicanonical weight factor for REMUCA and MUCAREM 
%which realize the random walk in energy space but not 
%get the enough flat probability distributions of energy to obtain 
%accurate thermodynamic quantities. 
%In this case we need to refine these multicanonical weight factors. 
%Thus we recalculate the density of state by using the 
%reweighting techniques and the result of REMUCA or MUCARE production run.  
%And the new multicanonical weight factor is refined by 
%the obtained density of state.
%This is why the random walks of energy space in 
%REMUCA and MUCAREM simulations are more frequent than that of REM 
%and then the density of states obtained by REMUCA and MUCAREM production run
%are more accurate than that by only REM simulation. 
%The weight factor is obtained by this density of state and 
%become extremely precision. 
%The multicanonical production runs with the refined weight factors 
%are more frequent random walks in energy space and 
%can obtain more flat probability distributions of energy 
%to accurate thermodynamic quantities. 

To be more specific, this iterative process can be 
summarized as follows.
%We brifly describe the formulations of applications of REMUCA and MUCAREM 
%as follows. 
The REMUCA production run corresponds to 
a MUCA simulation with the weight factor $W_{\rm mu}(E)$. 
%(see Eq. (\ref{eqn19}) in Paper I). 
%(see Eq. (25) in Paper I). 
The new estimate of the density of states can be obtained
by the single-histogram reweighting techniques 
of Eq. (\ref{eqn10c}). 
%as follows. 
%(see Eq. (\ref{eqn2c}) in Paper I):
%(see Eq. (5) in Paper I):
%\begin{equation} 
%n(E)= \displaystyle{\frac{N_{\rm mu}(E)}{W_{\rm mu}(E)}}~,
%\label{eqn103}
%\end{equation}
%where $N_{\rm mu} (E)$ is the energy histogram obtained from the 
%MUCA production run in REMUCA.
%$W_{RM}$(E) is the multicanocanil weight factor 
%in REMUCA (see Eq.(\label{eqn19}) 
%By substituting this density of state of Eq. (\ref{eqn2c}) 
%into Eqs. (\ref{eqn19}) and (\ref{eqn20}) in paper I,
%one can get the more accurate weight factor 
%for multicanonical simulation. 
On the other hand, from the MUCAREM production run, 
the improved density of states can be obtained
by the multiple-histogram reweighting
techniques of Eqs. (\ref{eqn30}) and (\ref{Eqn31}).
%(see Eqs. (\ref{eqn15}) and (\ref{eqn16}) in Paper I):
%(see Eqs. (40) and (41) in Paper I):
%\begin{equation}
%%n(E) = \frac{  \displaystyle{ \sum_{m=1}^{{\cal M}} g_m^{-1} N_{\rm mu}^{\{m\}} (E)} }
%%{ \displaystyle{ \sum_{m=1}^{{\cal M}} g_m^{-1} n_m \exp(f_m ) W_{\rm mu}^{\{m\}}(E)} },
%n(E) = \frac{  \displaystyle{ \sum_{m=1}^{{\cal M}} N_{\rm mu}^{\{m\}} (E)} }
%{ \displaystyle{ \sum_{m=1}^{{\cal M}} n_m \exp(f_m ) W_{\rm mu}^{\{m\}}(E)} }~,
%\label{eqn104}
%\end{equation}
%and 
%\begin{equation}
%\exp(-f_m) \equiv \sum_E n(E) W_{\rm mu}^{\{m\}}(E)~.
%\label{eqn105}
%\end{equation}
%Here, $N_{\rm mu}^{\{m\}}(E)$ and $n_m$ 
%are the energy histogram and the total number of 
%samples obtained for the $m$-th multicanonical ensemble
%(or, energy range) in the MUCAREM production run, respectively,
%and $W_{\rm mu}^{\{m\}}(E)$ is the corresponding
%multicanonical weight factor .
%%(see Eq. (34) in Paper I).
%Here, $g_m$, $n(E)$, and $f_m$ are obtained by the process as in Paper I. 

The improved density of states thus obtained 
leads to a new multicanonical
weight factor (see Eq. (\ref{eqn6})).
The next iteration can be either a MUCA production run (as in REMUCA) or
MUCAREM production run.  The results of this production run may yield
an optimal multicanonical weight factor that yields
a sufficiently flat energy distribution for the entire
energy range of interest.  If not, we can repeat the 
above process by obtaining the third estimate of the multicanonical
weight factor either by a MUCA production run (as in REMUCA) or
by a MUCAREM production run, and so on.
%from Eq. (\ref{eqn103}) or from 
%Eq. (\ref{eqn104}), and so on.

We remark that as the estimate of the multicanonical weight factor becomes
more accurate, one is required to have a less number of
replicas for a successful MUCAREM simulation, because each replica
will have a flat energy distribution for a wider energy range.  Hence, 
for a large, complex system, it is often more efficient to
first try MUCAREM and iteratively reduce the number of replicas so
that eventually one needs only one or a few replicas (instead of trying
REMUCA directly from the beginning and iterating MUCA simulations). \\
   
%By substituting the density of state of Eq. (\ref{eqn15}) 
%into Eqs. (\ref{eqn19}) and (\ref{eqn20}) in paper I of the series,
%one can get the more accurate weight factor for multicanonical simulation. 
%Remark that 
%we also can get the refined weight factor of MUCAREM by substituting 
%the value of Eqs. (\ref{eqn24}) and (\ref{eqn25}) into 
%Eq. (in paper I of the series). 
%Owing to decrease the number of ${\cal M}$, the MUCAREM simulation with 
%small replica are realized. \\
%%%

We now describe the simulated tempering replica-exchange method
(STREM) \cite{STREM}.
Suppose that the simulated tempering weight factor $W_{ST}(E;T_n)$
(or equivalently, the dimensionless Helmholtz free energy
$a_{n}$ in Eq. (\ref{Eqn3}))
has been obtained as in REST or by any other methods
in the entire temperature range of interest ($T_1 \le T_n \le 
T_M$). 
We devide the overlapping temperature ranges into
${\cal M}$ regions (${\cal M} \ll M$).
Suppose each temperature range $m$ has ${\cal N}_m$ temperatures:
$T_k^{\{m\}}$ ($k=1,\cdots,{\cal N}_m$) for
$m=1,\cdots,{\cal M}$.
We assign each temperature range to a replica;
each replica $i$
is in one-to-one correspondence with a
different temperature range $m$ of ST run, where
$T_1^{\{m\}} \le T_k^{\{m\}} \le T_{{\cal N}_m}^{\{m\}}$
($k=1,\cdots,{\cal N}_m$).
We then introduce the replica-exchange process
between neighboring temperature ranges.
This works when we allow sufficient overlaps between
the temperature regions.

A STREM simulation is then realized by alternately 
performing the following two steps. \cite{STREM}
\begin{enumerate}
\item Each replica 
performs a ST simulation within the fixed temperature range
$simultaneously$ and $independently$ for a certain MC or MD steps.
\item A pair of replicas, say $i$ and $j$, which are at, say
$T=T_k^{\{m\}}$ and $T=T_{\ell}^{\{m+1\}}$, in neighboring 
temperature ranges, say $m$-th and $(m+1)$-th, 
respectively, are exchanged:
$X = \left\{\cdots, x_k^{[i]}, \cdots, x_{\ell}^{[j]}, \cdots \right\}
\longrightarrow \
X^{\prime} = \left\{\cdots, x_k^{[j]}, \cdots, x_{\ell}^{[i]},
\cdots \right\}$.
The transition probability of this replica exchange is 
given by the Metropolis criterion:
\begin{equation}
w(X \rightarrow X^{\prime})
= {\rm min}\left(1,\exp \left( - \Delta \right)\right)~,
%= \left\{
%\begin{array}{ll}
% 1~, & {\rm for} \ \Delta \le 0~, \cr
% \exp \left( - \Delta \right)~, & {\rm for} \ \Delta > 0~,
%\end{array}
%\right.
\label{Eqn22}
\end{equation}
where
\begin{equation}
\Delta \equiv \left(\beta_k^{\{m\}} - \beta_{\ell}^{\{m+1\}} \right)
              \left(E\left(q^{[j]}\right)
                  - E\left(q^{[i]}\right)\right)~. 
\label{Eqn23}
\end{equation}

\end{enumerate}

While in MUCAREM each replica performs a random walk
in multicanonical ensemble of finite energy range,
in STREM each replica performs a random walk by
simulated tempering of finite temperature range.
These ``local'' random walks are made ``global'' to cover the
entire energy range of interest by the replica-exchange
process.

\subsection{Multidimensional Replica-Exchange Method}

We now present our multidimensional extension of REM, which
we refer to as {\it multidimensional replica-exchange method}
(MREM) \cite{SKO}.
The crucial observation that led to the new algorithm is:  
As long as we have $M$ {\it non-interacting}
replicas of the original system, the Hamiltonian 
$H(q,p)$ of the system does not have to be identical
among the replicas and it can depend on a parameter
with different parameter values for different replicas.
Namely, we can write the Hamiltonian for the $i$-th
replica at temperature $T_m$ as
\begin{equation}
H_m (q^{[i]},p^{[i]}) =  K(p^{[i]}) + E_{\lambda_m} (q^{[i]})~,
\label{Eqn16}
\end{equation}
where the potential energy $E_{\lambda_m}$ depends on a
parameter $\lambda_m$ and can be written as
\begin{equation}
E_{\lambda_m} (q^{[i]}) = E_0 (q^{[i]}) + \lambda_m V(q^{[i]})~.
\label{Eqn16p}
\end{equation}
This expression for the potential energy is often used in
simulations.
For instance, in umbrella sampling \cite{US}, $E_0(q)$ and
$V(q)$ can be respectively taken as the original potential
energy and the ``biasing'' potential energy with the
coupling parameter $\lambda_m$.  In simulations of spin 
systems, on the other hand, 
$E_0(q)$ and $V(q)$ (here, $q$ stands for spins)
can be respectively considered as the
zero-field term and the magnetization term coupled with
the external field $\lambda_m$. 

While replica $i$ and temperature
$T_m$ are in one-to-one correspondence
in the original REM,
replica $i$ and ``parameter set''
$\Lambda_m \equiv (T_m,\lambda_m)$ are in one-to-one
correspondence in the new algorithm.
Hence, the present algorithm can be considered as a
multidimensional extension of the original replica-exchange
method where the ``parameter space'' is one-dimensional 
(i.e., $\Lambda_m = T_m$).
Because the replicas are non-interacting, the weight factor 
for the state $X$ in this new
generalized ensemble is again given by the product
of Boltzmann factors for each replica (see Eq.~(\ref{eq7})):
\begin{equation}
\begin{array}{rl}
W_{\rm MREM}(X) &= \exp \left\{- \dis{\sum_{i=1}^M \beta_{m(i)} 
H_{m(i)}\left(q^{[i]},p^{[i]}\right) } \right\}~, \cr
 &= \exp \left\{- \dis{\sum_{m=1}^M \beta_m 
H_m\left(q^{[i(m)]},p^{[i(m)]}\right) }
 \right\}~,
\end{array}
\label{Eqn19}
\end{equation}
where $i(m)$ and $m(i)$ are the permutation functions in 
Eq.~(\ref{eq4b}).
Then the same derivation
that led to the original replica-exchange
criterion follows, and the
transition probability of replica exchange is
given by Eq.~(\ref{eq15}), where
we now have (see Eq.~(\ref{eqn14a})) \cite{SKO}
\begin{equation}
\Delta = \beta_m 
\left(E_{\lambda_m}\left(q^{[j]}\right) - 
E_{\lambda_m}\left(q^{[i]}\right)\right) 
- \beta_n
\left(E_{\lambda_n}\left(q^{[j]}\right) - 
E_{\lambda_n}\left(q^{[i]}\right)\right)~.
\label{eqn21}
\end{equation}
Here, $E_{\lambda_m}$ and $E_{\lambda_n}$ are the
total potential energies (see Eq.~(\ref{Eqn16p})).
Note that we need to newly evaluate the potential
energy for exchanged coordinates,
$E_{\lambda_m} (q^{[j]})$ and $E_{\lambda_n} (q^{[i]})$,
because $E_{\lambda_m}$ and $E_{\lambda_n}$ are in general
different functions.

For obtaining the canonical distributions,
the multiple-histogram reweighting techniques \cite{FS2,WHAM}
are particularly suitable.
Suppose we have made a single run of the present
replica-exchange simulation with $M$ replicas that correspond
to $M$ different parameter sets
$\Lambda_m \equiv (T_m,\lambda_m)$ ($m=1, \cdots, M$).
Let $N_m(E_0,V)$ and $n_m$
be respectively 
the potential-energy histogram and the total number of
samples obtained for the $m$-th parameter set
$\Lambda_m$.
The WHAM equations that yield the canonical
probability distribution 
$P_{T,\lambda} (E_0,V)=n(E_0,V)\exp(-\beta E_\lambda)$
with any
potential-energy parameter value $\lambda$ at
any temperature $T=1/k_{\rm B} \beta$
are then given by \cite{SKO}
\begin{equation}
%P_{T,\lambda} (E_0,V)
%= \left[\frac{\dis{\sum_{m=1}^M g_m^{-1}~N_m(E_0,V)}} 
%{\dis{\sum_{m=1}^M g_m^{-1}~n_{m}~e^{f_m-\beta_m E_{\lambda_m}}}} \right]
%e^{-\beta E_{\lambda}}~,
n(E_0,V)
= \frac{\dis{\sum_{m=1}^M g_m^{-1}~N_m(E_0,V)}} 
%{\dis{\sum_{m=1}^M g_m^{-1}~n_{m}~e^{f_m-\beta_m E_{\lambda_m}}}}~,
{\dis{\sum_{m=1}^M g_m^{-1}~n_{m}~\exp \left(f_m-\beta_m E_{\lambda_m}\right)}}~,
\label{eqn19}
\end{equation}
and for each $m$ ($=1, \cdots, M$)
\begin{equation}
%e^{-f_m} = \sum_{E_0,V} n(E_0,V) e^{-\beta_m E_{\lambda_m}}~.
\exp (-f_m) = \sum_{E_0,V} n(E_0,V) \exp \left(-\beta_m E_{\lambda_m}\right)~.
\label{eqn20}
\end{equation}
Here, $n(E_0,V)$ is the generalized density of states.
Note that $n(E_0,V)$ is independent of the parameter sets
$\Lambda_m \equiv (T_m,\lambda_m)$ ($m=1, \cdots, M$).
The density of states
$n(E_0,V)$ and the ``dimensionless''
Helmholtz free energy $f_m$ in Eqs.~(\ref{eqn19}) and
(\ref{eqn20}) are solved self-consistently
by iteration.
% \cite{FS2,WHAM}. 

Incidentally, these formulations of MREM give multidimensional
extensions of REMUCA \cite{SO3,MSO03} and REST \cite{MO4}.
In the former, we obtain uniform distributions both in $E_0$
and $V$, whereas in the latter, the parameter sets $\Lambda_m$
become dynamical variables and a uniform distribution in
those parameters will be obtained.  Namely, after a short
MREM simulation, we can use the multiple-histogram reweighting
techniques of Eqs. (\ref{eqn19}) and (\ref{eqn20}) to obtain
$n(E_0,V)$ and $f_m$.  Hence, we can determine
the multidimensional multicanonical weight factor $W_{\rm mu}(E_0,V)$
and the multidimensional simulated tempering weight factor
$W_{\rm ST}(E_0,V;\Lambda_m)$.  The former is given by
\begin{equation}
W_{\rm mu}(E_0,V) = \frac{1}{n(E_0,V)}~,
\label{eqn20b}
\end{equation}
and the latter is given by (see Eq. (\ref{Eqn3}))
\begin{equation}
%W_{ST}(E_0,V;\Lambda_m) = e^{-\beta_m E_{\lambda_m} + f_m}~.
W_{\rm ST}(E_0,V;\Lambda_m) = \exp \left(-\beta_m E_{\lambda_m} + f_m \right)~.
\label{eqn20c}
\end{equation}

We can use MREM for
free energy calculations.  We first describe the free-energy
perturbation case.  The potential energy is given by
\begin{equation}
E_{\lambda} (q) = E_I (q) + \lambda \left(E_F (q) - E_I (q)\right)~,
\label{eqn22}
\end{equation}
where $E_I$ and $E_F$ are the potential energy for
a ``wild-type'' molecule and a ``mutated''
molecule, respectively.  Note that this equation has the same
form as Eq.~(\ref{Eqn16p}).

Our replica-exchange simulation is performed for $M$ replicas
with $M$ different values of the parameters
$\Lambda_m = (T_m,\lambda_m)$.
Since $E_{\lambda = 0} (q) = E_I (q)$ and
$E_{\lambda = 1} (q) = E_F (q)$, we should choose enough
$\lambda_m$ values distributed in the range between 0 and 1
so that we may have sufficient acceptance of replica exchange.
From the simulation, $M$ histograms $N_m (E_I,E_F-E_I)$, or
equivalently $N_m(E_I,E_F)$, are obtained.  The Helmholtz
free energy difference of ``mutation'' at 
temperature $T$ $(=1/k_{\rm B}\beta)$,
$\Delta F \equiv F_{\lambda = 1} - F_{\lambda = 0}$, can then
be calculated from 
\begin{equation}
\exp (-\beta \Delta F) = \frac{Z_{T,\lambda=1}} 
{Z_{T,\lambda=0}} =  
\frac{\dis{\sum_{E_I,E_F}
P_{T,\lambda=1} (E_I,E_F)}}
{\dis{\sum_{E_I,E_F}
P_{T,\lambda=0} (E_I,E_F)}} ~,
\label{Eqn24}
\end{equation}
%where $P_{T,\lambda} (E_I,E_F) = n(E_I,E_F) e^{-\beta E_{\lambda}}$ 
where $P_{T,\lambda} (E_I,E_F) = n(E_I,E_F) \exp \left(-\beta E_{\lambda}\right)$ 
are obtained from the WHAM
equations of Eqs.~(\ref{eqn19}) and (\ref{eqn20}).

We now describe another free energy calculations based on
MREM applied to umbrella sampling \cite{US},
which we refer to as 
{\it replica-exchange umbrella sampling} (REUS).
The potential energy is a generalization of Eq.~(\ref{Eqn16p})
and is given by
\begin{equation}
E_{\bil} (q) = E_0 (q) + \sum_{\ell = 1}^L
\lambda^{(\ell)} V_{\ell} (q)~,
\label{Eqn25}
\end{equation}
where $E_0(q)$ is the original unbiased potential, 
$V_{\ell}(q)$ ($\ell =1, \cdots, L$) are the 
biasing (umbrella) potentials, and $\lambda^{(\ell)}$ are the
corresponding coupling constants
($\bil = (\lambda^{(1)}, \cdots, \lambda^{(L)})$).
Introducing a ``reaction coordinate'' $\xi$,
the umbrella potentials are usually written as harmonic
restraints:
\begin{equation}
V_{\ell} (q) = k_{\ell} \left( \xi (q) - d_{\ell} \right)^2~,
~(\ell =1, \cdots, L)~,
\label{Eqn26}
\end{equation}
where $d_{\ell}$ are the midpoints and $k_{\ell}$ are the
strengths of the restraining potentials.
We prepare $M$ replicas with $M$
different values of the parameters
$\biL_m = (T_m,\bil_m)$, and the replica-exchange
simulation is performed.  Since the umbrella potentials
$V_{\ell} (q)$ in Eq.~(\ref{Eqn26})
are all functions of the reaction coordinate
$\xi$ only, we can take the histogram
$N_m (E_0,\xi)$ instead of
$N_m (E_0,V_1, \cdots, V_L)$.
The WHAM equations of
Eqs.~(\ref{eqn19}) and (\ref{eqn20}) can then be written as \cite{SKO}
\begin{equation}
%P_{T,\bil} (E_0,\xi)
%= \left[\frac{\dis{\sum_{m=1}^M g_m^{-1}~N_m(E_0,\xi)}} 
%{\dis{\sum_{m=1}^M g_m^{-1}~n_{m}~e^{f_m-\beta_m E_{\bil_m}}}} \right]
%e^{-\beta E_{\bil}}~,
n(E_0,\xi)
= \frac{\dis{\sum_{m=1}^M g_m^{-1}~N_m(E_0,\xi)}} 
%{\dis{\sum_{m=1}^M g_m^{-1}~n_{m}~e^{f_m-\beta_m E_{\bil_m}}}}~
{\dis{\sum_{m=1}^M g_m^{-1}~n_{m}~\exp \left(f_m-\beta_m E_{\bil_m}\right)}}~
\label{Eqn27}
\end{equation}
and for each $m$ ($=1, \cdots, M$)
\begin{equation}
%%e^{-f_m} = \sum_{E_0,\xi} P_{T_m,\bil_m} (E_0,\xi)~.
%e^{-f_m} = \sum_{E_0,\xi} n(E_0,\xi) e^{-\beta_m E_{\bil_m}}~.
\exp (-f_m) = \sum_{E_0,\xi} n(E_0,\xi) \exp \left(-\beta_m E_{\bil_m}\right)~.
\label{Eqn28}
\end{equation}
The expectation value of a physical quantity $A$ 
with any
potential-energy parameter value $\bil$ at
any temperature $T$ ($=1/k_{\rm B} \beta$) is now
given by
\begin{equation}
<A>_{T,\bil} \ = \frac{\dis{\sum_{E_0,\xi}
A(E_0,\xi) P_{T,\bil} (E_0,\xi)}}
{\dis{\sum_{E_0,\xi} P_{T,\bil} (E_0,\xi)}}~,
\label{Eqn29}
\end{equation}
%where $P_{T,\bil} (E_0,\xi) = n(E_0,\xi) e^{-\beta E_{\bil}}$
where $P_{T,\bil} (E_0,\xi) = n(E_0,\xi) \exp \left(-\beta E_{\bil}\right)$
is obtained from the WHAM
equations of Eqs.~(\ref{Eqn27}) and (\ref{Eqn28}).

The potential of mean force (PMF), or free energy as a function of
the reaction coordinate, of the original, unbiased system 
at temperature $T$ is given by
\begin{equation}
{\cal W}_{T,\bil = \{0\}} (\xi) = - k_{\rm B} T \ln
\left[ \sum_{E_0} P_{T,\bil = \{0\}} (E_0,\xi) \right]~,
\label{Eqn30}
\end{equation}
where $\{0\} = (0, \cdots, 0)$.

We now present two examples of realization of REUS.
In the first example, we use only one temperature, $T$, and 
$L$ umbrella potentials.
%Let us order the umbrella potentials, $V_{\ell}$ in Eq.~(\ref{Eqn25}),
%in the increasing order of the midpoint value $d_{\ell}$,
%i.e., the same order that appears in Table 1.
We prepare replicas so that the potential energy for each
replica includes exactly one umbrella potential
(here, we have $M = L$).
Namely, in Eq.~(\ref{Eqn25}) for $\bil = \bil_m$ we set
\begin{equation}
\lambda^{(\ell)}_m = \delta_{\ell,m}~,
\label{Eqnn31}
\end{equation}
where $\delta_{k,l}$ is Kronecker's delta function, and
we have
\begin{equation}
E_{\bil_m} (q^{[i]}) = E_0 (q^{[i]}) + V_m (q^{[i]})~.
\label{Eqn32}
\end{equation}
%The difference between REUS1 and US1 is whether replica exchange
%is performed or not during the parallel MD simulations.
%In REUS1 seven pairs of
We exchange 
replicas corresponding to ``neighboring'' umbrella potentials,
$V_{m}$ and $V_{m+1}$.
%were simultaneously exchanged after every 200 fs of parallel
%MD simulations, and the pairing was alternated between
%the two possible choices.  (Other pairings will have much smaller 
%acceptance ratios of replica exchange.)
The acceptance criterion for replica exchange is given
by Eq.~(\ref{eq15}), where Eq.~(\ref{eqn21}) now reads
(with the fixed inverse temperature $\beta = 1/k_{\rm B} T$) \cite{SKO}
\begin{equation}
\Delta = \beta 
\left(V_m\left(q^{[j]}\right) - 
      V_m\left(q^{[i]}\right) -
      V_{m+1}\left(q^{[j]}\right) + 
      V_{m+1}\left(q^{[i]}\right)\right)~,
\label{Eqn33}
\end{equation}
where replica $i$ and $j$ respectively have umbrella potentials
$V_m$ and $V_{m+1}$ before the exchange.

In the second example, we prepare $N_T$ temperatures
and $L$ umbrella potentials,
which makes the total
%In REUS2 and US2, 16 replicas were simulated at four different
%temperatures with four different restraining potentials
%(there are $L=4$
%umbrella potentials at $N_T=4$ temperatures, making the total
number of replicas $M=N_T \times L$.
We can introduce the following re-labeling for the parameters
that characterize the replicas:
\begin{equation}
\begin{array}{rl}
\biL_m = (T_m,\bil_m) & \longrightarrow
\ \biL_{I,J} = (T_I,\bil_J)~. \cr
(m=1, \cdots, M) & \ \ \ \ \ \ \ \ \ \ (I=1, \cdots, N_T,~J=1, \cdots, L)
\end{array}
\label{Eqn34}
\end{equation}
The potential energy is given by Eq.~(\ref{Eqn32})
with the replacement: $m \rightarrow J$.
%Let us again order the umbrella potentials, $V_J$,
%and the temperatures, $T_I$, in
%the same order that appear in Table 1.
%The difference between REUS2 and US2 is 
%whether replica exchange
%is performed or not during the MD simulations.
%In REUS2 we performed the following replica-exchange processes alternately
%after every 200 fs of parallel MD simulations:
We perform the following replica-exchange processes alternately:
\begin{enumerate}
\item Exchange pairs of replicas corresponding to neighboring temperatures,
$T_I$ and $T_{I+1}$ 
(i.e., exchange replicas $i$ and $j$ that
respectively correspond to parameters
$\biL_{I,J}$ and $\biL_{I+1,J}$).
(We refer to this process as $T$-exchange.)
\item Exchange pairs of replicas corresponding to 
``neighboring'' umbrella potentials,
$V_J$ and $V_{J+1}$ 
(i.e., exchange replicas $i$ and $j$ that
respectively correspond to parameters
$\biL_{I,J}$ and $\biL_{I,J+1}$).
(We refer to this process as $\lambda$-exchange.)
\end{enumerate}
%In each of the above processes, two pairs of replicas were simultaneously
%exchanged, and the pairing was further 
%alternated between the two possibilities.
The acceptance criterion for these 
replica exchanges is given by 
Eq.~(\ref{eq15}), where Eq.~(\ref{eqn21}) now reads \cite{SKO}
\begin{equation}
\Delta = \left(\beta_{I} - \beta_{I+1} \right)
\left(E_0 \left(q^{[j]}\right) 
    + V_J \left(q^{[j]}\right) 
    - E_0 \left(q^{[i]}\right)
    - V_J \left(q^{[i]}\right)\right)~, 
\label{Eqn35}
\end{equation}
for $T$-exchange, and
\begin{equation}
\Delta = \beta_I 
\left(V_J\left(q^{[j]}\right) - 
      V_J\left(q^{[i]}\right) -
      V_{J+1}\left(q^{[j]}\right) + 
      V_{J+1}\left(q^{[i]}\right)\right)~,
\label{Eqn36}
\end{equation}
for $\lambda$-exchange.
By this procedure, the random walk 
in the reaction coordinate space as well as in the temperature
space can be realized.

%\noindent
%{\bf CONCLUSIONS} \\
\section{CONCLUSIONS}

In this article we have reviewed uses of generalized-ensemble
algorithms for both Monte Carlo simulations and molecular dynamics
simulations.
A simulation in generalized ensemble realizes a random
walk in potential energy space, alleviating the
multiple-minima problem that is a common difficulty
in simulations of complex systems with many degrees
of freedom.

Detailed formulations of the three well-known
generalized-ensemble algorithms, namely,
multicaonical algorithm (MUCA),
simulated tempering (ST), and replica-exchange method (REM), 
were given.  
%Among the three, MUCA and ST are closely
%related (actually, their weights can be converted into
%each other by Laplace transformations), while REM
%introduces a new ingredient that it is simulated with
%multiple copies (or replicas) of the system.
%The weight factor for MUCA is (the inverse of) the
%density of states, and that for ST is the (dimensionless)
%Helmholtz free energy.  These
%weight factors are not {\it a priori} known and have to be
%determined by iterations of short trial simulations.
%This process can be non-trivial and very tedius for
%complex systems with many degreees of freedom, whereas
%the weight factor of REM is simply
%a product of Boltzmann factors and known.
%However, REM also has a computational difficulty:
%As the number of degrees of freedom of the system increases,
%the required number of replicas also greatly increases, whereas 
%only a single replica is simulated in MUCA or ST.
%This demands a lot of computer power for complex systems.

We then introduced five new generalized-ensemble
algorithms that combine the merits of the above three methods. 
We refer to these methods as replica-exchange multicanonical
algorithm (REMUCA), replica-exchange simulated tempering (REST),
multicanonical replica-exchange method (MUCAREM), simulated
tempering replica-exchange method (STREM), and multidimensional
replica-exchange method (MREM), the last of which also
led to replica-exchange umbrella sampling (REUS).

The question is then which method is the most recommended.
We have recently studied the effectiveness of MUCA, REM,
REMUCA, and MUCAREM in the protein folding problem \cite{MSO03}.  
Our criterion for the
effectiveness was how many times the random walk cycles
between the high-energy region and low-energy region 
are realized within a fixed 
number of total MC (or MD) steps.
We found that once the optimal MUCA weight factor is
obtained, MUCA (and REMUCA) is the most effective
(i.e., has the most number of
random walk cycles), and REM is the least \cite{MSO03}.
We also found that once the optimal ST weight factor
is obtained, ST (and REST) has more random walk cycles
than REM \cite{MO4,STREM}.  Moreover, we compared
the efficiency of  
Berg's recursion \cite{MUCAW},
Wang-Landau method \cite{Landau1,Landau2},
and REMUCA/MUCAREM as methods for the multicanonical weight factor
determination in two-dimensional 10-state Potts model
and found that the three methods are about equal in efficiency
\cite{NSMO}--\cite{O03}.
%\cite{NSMO,NSMO02,O03}.

Hence, the answer to the above question will depend on 
how much time one is willing to (or forced to)
spend in order to determine the MUCA or ST weight factors.
Given a problem, the first choice is REM because of its
simplicity (no weight factor determination is required).
If REM turns out to be insufficient or too much 
time-consuming (like the case with
first-order phase transitions), then other
more powerful algorithms such as those
presented in the present article are recommended.

\vspace{0.5cm}
\noindent
{\bf Acknowledgements}: \\
The author would like to thank his co-workers for useful
discussions.  In particular, he is grateful to Drs. B.A. Berg, 
U.H.E. Hansmann, A. Mitsutake, and Y. Sugita for 
collaborations that led to the new generalized-ensemble algorithms
described in the present article.
This work was supported, in part, by grants from 
the Research for the Future Program of the Japan Society for the 
Promotion of Science (JSPS-RFTF98P01101) and
the NAREGI Nanoscience Project, Ministry of
Education, Culture, Sports, Science and Technology, Japan.\\

%%%%%%%%%%%%%%%%%%%%%%%%% references %%%%%%%%%%%%%%%%%%%


\noindent
\begin{thebibliography}{(00)}
\bibitem{Metro} Metropolis, N., Rosenbluth, A.W., Rosenbluth, M.N.,
Teller, A.H., and Teller, E. (1953) {\it J. Chem. Phys.} {\bf 21},
1087--1092.

\bibitem{RevSch} V{\'a}squez, M., N{\'e}methy, G., and 
Scheraga, H.A. (1994)
{\it Chem. Rev.} {\bf 94}, 2183--2239.
\bibitem{BeSt} Berne, B.J. and Straub, J.E. (1997)
{\it Curr. Opin. Struct. Biol.} {\bf 7}, 181--189.
\bibitem{RevHO2} Hansmann, U.H.E. and Okamoto, Y. (1999)
{\it Curr. Opin. Struct. Biol.} {\bf 9}, 177--183.

%\bibitem{RevO} Okamoto, Y. (1998) {\it Recent Res. Devel. in Pure} and
%        {\it Applied Chem.} {\bf 2}, 1--23.
\bibitem{RevHO1} Hansmann, U.H.E. and Okamoto, Y. (1999)
in {\it Annual Reviews of Computational Physics VI}, Stauffer, D., Ed.,
World Scientific, Singapore, pp. 129--157.
\bibitem{RevMSO} Mitsutake, A., Sugita, Y., and Okamoto, Y. (2001) 
{\it Biopolymers (Peptide Science)} {\bf 60}, 96--123.
\bibitem{RevSO} Sugita, Y. and Okamoto, Y. (2002)
in {\it Lecture Notes in Computational Science and Engineering}, 
Schlick, T. and Gan, H.H., Eds.,
Springer-Verlag, Berlin, pp. 304--332; cond-mat/0102296.

\bibitem{Be02} Berg, B.A. (2002) 
% {\it Generalized Ensemble Simulations for Complex Systems}, 
{\it Comp. Phys. Commun.} {\bf 147}, 
%52.
52--57.


\bibitem{FS1} Ferrenberg, A.M. and Swendsen, R.H. (1988)
{\it Phys. Rev. Lett.} {\bf 61}, 2635--2638; {\it ibid.}
(1989) {\bf 63}, 1658.
\bibitem{FS2} Ferrenberg, A.M. and Swendsen, R.H. (1989)
{\it Phys.  Rev.  Lett.} {\bf 63}, 1195--1198.
\bibitem{WHAM} Kumar, S., Bouzida, D., Swendsen, R.H., Kollman, P.A.,
and Rosenberg, J.M. (1992) {\it J. Comput. Chem.}
  {\bf 13}, 1011--1021.
  
\bibitem{SW87} Swendsen, R.H. and Wang, J.S. (1987)
{\it Phys.  Rev.  Lett.} {\bf 58}, 86--88.

\bibitem{Wolff89} Wolff, U. (1989) {\it Phys. Rev. Lett.} 
{\bf 62}, 361--364.
\bibitem{ELM93} Evertz, H.G., Lana, G., and Marcu, M. (1993) 
{\it Phys. Rev. Lett.} {\bf 70}, 875--879.

%\bibitem{KSWF03} Klepeis, J.L., Schafroth, H.D., Westerberg, K.M.,
%and Floudas, C.A. (2002) 
%{\it Adv. Chem. Phys.} {\bf 120}, 265--457.


\bibitem{MUCA1} Berg, B.A. and Neuhaus, T. (1991) {\it Phys. Lett.} {\bf 
B267},
  249--253.
\bibitem{MUCA2} Berg, B.A. and Neuhaus, T. (1992) {\it Phys. Rev. Lett.} 
{\bf 68}, 9--12.
\bibitem{MUCArev} Berg, B.A. (2000) 
{\it Fields Institute Communications} {\bf 26}, 1--24; also see
cond-mat/9909236.

%\bibitem{RevJanke} Janke, W. (1998) 
%{\it Physica A} {\bf 254}, 164--178.


\bibitem{Lee} Lee, J. (1993) {\it Phys. Rev. Lett.} {\bf 71}, 211--214;
 {\it ibid.} {\bf 71}, 2353.
\bibitem{MZ} Mezei, M. (1987)
 {\it J. Comput. Phys.} {\bf 68}, 237--248.
\bibitem{BK} Bartels, C. and Karplus, M. (1998)
 {\it J. Phys. Chem. B} {\bf 102}, 865--880.
\bibitem{Landau1} Wang, F. and Landau, D.P. (2001)
{\it Phys. Rev. Lett.} {\bf 86}, 2050--2053.
\bibitem{Landau2} Wang, F. and Landau, D.P. (2001)
{\it Phys. Rev. E} {\bf 64}, 056101.
   
\bibitem{dePablo} Yan, Q., Faller, R., and de Pablo, J.J. (2002)
{\it J. Chem. Phys.} {\bf 116}, 8745--8749.

\bibitem{US} Torrie, G.M. and Valleau, J.P. (1977) 
{\it J. Comput. Phys.} {\bf 23}, 187--199.
 
\bibitem{WS02} Wang, J.S. and Swendsen, R.H. (2002)
{\it J. Stat. Phys.} {\bf 106}, 245--285.

%\bibitem{BHO} Berg, B.A., Hansmann, U.H.E., and Okamoto, Y. (1995) 
% {\it J. Phys. Chem.} {\bf 99}, 2236--2237.
\bibitem{MUCA3} Berg, B.A. and Celik, T. (1992) {\it Phys. Rev. Lett.} 
{\bf 69}, 2292--2295.
%\bibitem{MUCA4} Berg, B.A., Celik, T. and Hansmann, U.H.E. (1993) 
%{\it Europhys. Lett.} {\bf 22}, 63--68.
\bibitem{BeHaNe93} Berg, B.A., Hansmann, U.H.E., and Neuhaus, T. (1993)
Phys. Rev. B {\bf 47}, 497--500.
\bibitem{JK95} Janke, W. and Kappler, S. (1995) {\it Phys. Rev. Lett.} 
{\bf 74}, 212--215.
\bibitem{HS} Hesselbo, B. and Stinchcombe,\ R.B. (1995) {\it Phys. Rev. Lett.} {\bf 74}, 
2151--2155.
\bibitem{MUCA5} Berg, B.A. and Janke, W. (1998) {\it Phys. Rev. Lett.} 
{\bf 80}, 4771--4774.
\bibitem{MUCA6} Hatano, N. and Gubernatis, J.E. (2000)
{\it Prog. Theor. Phys. (Suppl.)} {\bf 138}, 442--447. 
\bibitem{BBJ00} Berg, B.A., Billoire, A., and Janke, W. (2000) 
{\it Phys. Rev. B} {\bf 61}, 12143--12150.

\bibitem{HO} Hansmann, U.H.E. and Okamoto, Y. (1993)
{\it J. Comput. Chem.} {\bf 14}, 1333--1338.
\bibitem{HO94} Hansmann, U.H.E. and Okamoto, Y. (1994)
 {\it Physica} {\bf A212}, 415--437.
\bibitem{HSch} Hao, M.H. and Scheraga, H.A. (1994)
        {\it J. Phys. Chem.} {\bf 98}, 4940--4948.
\bibitem{OH1} Okamoto, Y., Hansmann, U.H.E., and Nakazawa, T. (1995)
   {\it Chem. Lett.} {\bf 1995}, 391--392. 
\bibitem{OH} Okamoto, Y. and Hansmann, U.H.E. (1995)
   {\it J. Phys. Chem.} 
   {\bf 99}, 11276--11287.

\bibitem{KI95} Kidera, A. (1995) 
{\it Proc. Natl. Acad. Sci. U.S.A.} {\bf 92}, 9886--9889.

\bibitem{Wilding95} Wilding, N.B. (1995) {\it Phys. Rev. E} 
{\bf 52}, 602--611.

\bibitem{KGS} Kolinski, A., Galazka, W. and Skolnick, J. (1996)
{\it Proteins} {\bf 26}, 271--287.
\bibitem{UT} Urakami, N. and Takasu, M. (1996)
{\it J. Phys. Soc. Jpn.} {\bf 65}, 2694--2699.
\bibitem{KPV} Kumar, S., Payne, P., and V{\' a}squez, M. (1996) 
{\it J. Comput. Chem.} {\bf 17}, 1269--1275.

\bibitem{HOE96} Hansmann, U.H.E., Okamoto, Y., and Eisenmenger, F. (1996)
{\it Chem. Phys. Lett.} {\bf 259}, 321--330.

\bibitem{HO96a} Hansmann, U.H.E. and Okamoto, Y. (1996) {\it Phys. Rev. E} 
{\bf 54}, 5863--5865.

\bibitem{HO96b} Hansmann, U.H.E. and Okamoto, Y. (1997) 
{\it J. Comput. Chem.} {\bf 18}, 920--933.

\bibitem{NNK} Nakajima, N., Nakamura, H., and Kidera, A. (1997)
{\it J. Phys. Chem. B} {\bf 101}, 817--824.
%\bibitem{HE} Eisenmenger, F. and Hansmann, U.H.E. (1997)
%{\it J. Phys. Chem. B} {\bf 101}, 3304--3310.
\bibitem{BK2} Bartels, C. and Karplus, M. (1997)
 {\it J. Comput. Chem.} {\bf 18}, 1450--1462.

\bibitem{HNSKN} Higo, J., Nakajima, N., Shirai, H., Kidera, A., 
and Nakamura, H.
(1997) {\it J. Comput. Chem.} {\bf 18}, 2086--2092.

\bibitem{NakaHKN} Nakajima, N., Higo, Kidera, A., and Nakamura, H.
(1997) {\it J. Chem. Phys.} {\bf 278}, 297--301.

\bibitem{NY} Noguchi, H. and Yoshikawa, K. (1997)
{\it Chem. Phys. Lett.} {\bf 278}, 184--188.
\bibitem{KGS2} Kolinski, A., Galazka, W., and Skolnick, J.
(1998) {\it J. Chem. Phys.} {\bf 108}, 2608--2617.
\bibitem{ICK} Iba, Y., Chikenji, G., and Kikuchi, M. 
(1998) {\it J. Phys. Soc. Jpn.}
{\bf 67}, 3327--3330.
\bibitem{N} Nakajima, N. (1998) {\it Chem. Phys. Lett.} {\bf 288}, 319--326.
\bibitem{HSchp} Hao, M.H. and Scheraga, H.A. (1998) {\it J. Mol. Biol.} 
{\bf 277}, 973--983.
\bibitem{SNHKN} Shirai, H., Nakajima, N., Higo, J., Kidera, A., 
and Nakamura, H.
(1998) {\it J. Mol. Biol.} {\bf 278}, 481--496.

\bibitem{SBK} Schaefer, M., Bartels, C., and Karplus, M. (1998)
{\it J. Mol. Biol.} {\bf 284}, 835--848.
\bibitem{MHO} Mitsutake, A., Hansmann, U.H.E., and Okamoto, Y. (1998)
{\it J. Mol. Graphics Mod.} {\bf 16}, 226--238; 262--263.
\bibitem{HO99a} Hansmann, U.H.E. and Okamoto, Y. (1999)
{\it J. Chem. Phys.} {\bf 110}, 1267--1276.

\bibitem{HO99} Hansmann, U.H.E. and Okamoto, Y. (1999)
{\it J. Phys. Chem. B} {\bf 103}, 1595--1604.
\bibitem{SUYH} Shimizu, H., Uehara, K., Yamamoto, K., and Hiwatari, Y.
(1999) {\it Mol. Sim.} {\bf 22}, 285--301.
\bibitem{ONHN} Ono, S., Nakajima, N., Higo, J., and Nakamura, H. (1999)
{\it Chem. Phys. Lett.} {\bf 312}, 247--254.
\bibitem{MO2} Mitsutake, A. and Okamoto, Y. (2000)
{\it J. Chem. Phys.} {\bf 112}, 10638--10647.
\bibitem{SKGS00} Sayano, K., Kono, H., Gromiha, M.M., and Sarai, A. (2000)
{\it J. Comput. Chem.} {\bf 21}, 954--962.
\bibitem{YCBM} Yasar, F., Celik, T., Berg, B.A., and Meirovitch, H. (2000)
{\it J. Comput. Chem.} {\bf 21}, 1251--1261.
\bibitem{MO3} Mitsutake, A., Kinoshita, M., Okamoto, Y., and Hirata, F.
(2000) {\it Chem. Phys. Lett.} {\bf 329}, 295--303.
\bibitem{CK00} Chikenji, J. and Kikuchi, M. (2000)
{\it Proc. Natl. Acad. Sci. U.S.A.} {\bf 97}, 14273--14277.

\bibitem{KHN02} Kamiya, N., Higo, J., and Nakamura, H. (2002)
{\it Protein Sci.} {\bf 11}, 2297--2307.
\bibitem{JPS02} Jang, S.M., Pak, Y., and Shin, S.M. (2002)
{\it J. Chem. Phys.} {\bf 116}, 4782--4786.

\bibitem{RdeP02} Rathore, N. and de Pablo, J.J. (2002)
{\it J. Chem. Phys.} {\bf 116}, 7225--7230.

\bibitem{KFN03} Kim, J.G., Fukunishi, Y., and Nakamura, H. (2003) 
{\it Phys. Rev. E} {\bf 67}, 011105.

\bibitem{RKP03} Rathore, N., Knotts, T.A. IV, and de Pablo, J.J.
(2003)
{\it J. Chem. Phys.} {\bf 118}, 4285--4290.

\bibitem{TMK03} Terada, T., Matsuo, Y., and Kidera, A. (2003)
{\it J. Chem. Phys.} {\bf 118}, 4306--4311.

\bibitem{BNO03} Berg, B.A., Noguchi, H., and Okamoto, Y.
(2003) {\it Phys. Rev. E}, in press; cond-mat/0305055.

\bibitem{OO03} Okumura, H. and Okamoto, Y.
(2003) submitted for publication; cond-mat/0306144.

\bibitem{Muna} Munakata, T. and Oyama, S. (1996) {\it Phys. Rev. E}
{\bf 54}, 4394--4398. 


\bibitem{ST1} Lyubartsev, A.P., Martinovski, A.A., Shevkunov, S.V., and
Vorontsov-Velyaminov, P.N. (1992) {\it J. Chem. Phys.} {\bf 96}, 
1776--1783.
\bibitem{ST2} Marinari E. and Parisi, G. (1992) {\it Europhys. Lett.} 
{\bf 19}, 451--458.
\bibitem{STrev} Marinari, E., Parisi, G., and Ruiz-Lorenzo, J.J. (1998)
in {\it Spin Glasses and Random Fields}, Young, A.P., Ed.,
World Scientific, Singapore, pp. 59--98.
\bibitem{IRB1} Irb{\"a}ck, A. and Potthast, F. (1995) {\it J. Chem. Phys.}
{\bf 103}, 10298--10305.
\bibitem{IRB2} Irb{\"a}ck, A. and Sandelin, E. (1999) {\it J. Chem. Phys.}
{\bf 110}, 12256--12262.

\bibitem{SmBr} Smith, G.R. and Bruce, A.D. (1996)
{\it Phys. Rev. E} {\bf 53}, 6530--6543.
\bibitem{H97c} Hansmann, U.H.E. (1997)
{\it Phys. Rev. E} {\bf 56}, 6200--6203.
\bibitem{MUCAW} Berg, B.A. (1998)
{\it Nucl. Phys. B} (Proc. Suppl.) {\bf 63A-C}, 982--984.
\bibitem{Janke03} Janke, W. (2003) 
to appear in: Computer Simulations of Surfaces and Interfaces, 
NATO Advanced Study Institute, edited by 
Landau, D.P., Milchev, A., and D{\" u}nweg, B. 
(Kluwer, Dordrecht, 2003), in press.

%\bibitem{Tsa} Tsallis, C. (1988) {\it J. Stat. Phys.}
%{\bf 52}, 479--487.
%\bibitem{HO96d} Hansmann, U.H.E. and Okamoto, Y. (1997)
%{\it Phys. Rev. E} {\bf 56}, 2228--2233.
%\bibitem{HEO98} Hansmann, U.H.E., Eisenmenger, F. and Okamoto, Y. (1998)
%{\it Chem. Phys. Lett.} {\bf 297}, 374--382.
%\bibitem{HMO97} Hansmann, U.H.E., Masuya, M. and Okamoto, Y. (1997)
%{\it Proc. Natl. Acad. Sci. U.S.A.} {\bf 94}, 10652--10656.
%\bibitem{HOO} Hansmann, U.H.E., Okamoto, Y. and Onuchic, J.N. (1999)
%{\it Proteins} {\bf 34}, 472--483.
%\bibitem{Str2} Andricioaei, I. and Straub, J.E. (1997)
%{\it J. Chem. Phys.} {\bf 107}, 9117--9124.
%\bibitem{Muna2} Munakata, T. and Mitsuoka, S. (2000) {\it J. Phys. Soc.
% Jpn.}
%{\bf 69}, 92--96.
%\bibitem{SA} Kirkpatrick, S., Gelatt, C.D. Jr. and Vecchi, M.P.
%   (1983) {\it Science} {\bf 220}, 671--680. 
%\bibitem{STsal} Tsallis, C. and Stariolo, D.A. (1996)
% {\it Physica} {\bf A233}, 395--406.
%\bibitem{Str1} Andricioaei, I. and Straub, J.E. (1996)
%{\it Phys. Rev. E} {\bf 53}, R3055--R3058.
%\bibitem{H97b} Hansmann, U.H.E. (1998)
%{\it Physica A} {\bf 242}, 250--257.
%\bibitem{RevStr} Straub, J.E. and Andricioaei, I. (1999)
%{\it Braz. J. Phys.} {\bf 29}, 179--186.
%\bibitem{RevHO3} Hansmann, U.H.E. and Okamoto, Y. (1999)
%{\it Braz. J. Phys.} {\bf 29}, 187--198.

\bibitem{RE1} Hukushima, K. and  Nemoto, K. (1996)
{\it J. Phys. Soc. Jpn.} {\bf 65}, 1604--1608.
\bibitem{RE1b} Hukushima, K., Takayama, H., and Nemoto, K. (1996)
{\it Int. J. Mod. Phys. C} {\bf 7}, 337--344.

\bibitem{RE2} Geyer, C.J. (1991) in {\it Computing Science and Statistics:
 Proc. 23rd Symp. on the Interface}, Keramidas, E.M., Ed.,
 Interface Foundation, Fairfax Station, pp. 156--163.
\bibitem{RE3} Swendsen, R.H. and Wang, J.-S. (1986)
 {\it Phys. Rev. Lett.} {\bf 57}, 2607--2609.
\bibitem{KT} Kimura, K. and Taki, K. (1991) in {\it Proc. 13th IMACS World Cong. on Computation and Appl. Math. (IMACS '91)}, Vichnevetsky, R. and Miller, J.J.H.,
Eds., vol. 2, pp. 827--828.
\bibitem{JWK} Frantz, D.D., Freeman, D.L.,
and Doll, J.D. (1990) 
{\it J. Chem. Phys.} {\bf 93}, 2769--2784.
\bibitem{RE4} Tesi, M.C., van Rensburg, E.J.J., Orlandini, E., and
Whittington, S.G. (1996) {\it J. Stat. Phys.} {\bf 82}, 155--181.
\bibitem{IBArev} Iba, Y. (2001) 
{\it Int. J. Mod. Phys. C} {\bf 12}, 623--656.

\bibitem{H97} Hansmann, U.H.E. (1997)
{\it Chem. Phys. Lett.} {\bf 281}, 140--150. 

\bibitem{SO} Sugita, Y. and Okamoto, Y. (1999)
{\it Chem. Phys. Lett.} {\bf 314}, 141--151.

\bibitem{WD} Wu, M.G. and Deem, M.W. (1999)
{\it Mol. Phys.} {\bf 97}, 559--580. 

\bibitem{SKO} Sugita, Y., Kitao, A., and Okamoto, Y. (2000)
{\it J. Chem. Phys.} {\bf 113}, 6042--6051.

\bibitem{SO3} Sugita, Y. and Okamoto, Y. (2000)
{\it Chem. Phys. Lett.} {\bf 329}, 261--270.

\bibitem{MO4} Mitsutake, A. and Okamoto, Y. (2000) 
{\it Chem. Phys. Lett.} {\bf 332}, 131--138.


\bibitem{Kol} Gront, D., Kolinski, A., and Skolnick, J. (2000) 
{\it J. Chem. Phys.} {\bf 113}, 5065--5071.
\bibitem{Verk01} Verkhivker, G.M., Rejto, P.A., Bouzida, D., Arthurs, S.,
Colson, A.B., Freer, S.T., Gehlhaar, D.K., Larson, V., Luty, B.A.,
Marrone, T., and Rose, P.W. (2001)
{\it Chem. Phys. Lett.} {\bf 337}, 181--189.

\bibitem{FWT02} Fukunishi, H., Watanabe, O., and Takada, S.
(2002) {\it J. Chem. Phys.} {\bf 116}, 9058--9067.

\bibitem{MSO03} Mitsutake, A., Sugita, Y., and Okamoto, Y.
(2003) {\it J. Chem. Phys.} {\bf 118}, 6664--6675;
{\it ibid.} {\bf 118}, 6676--6688.

\bibitem{SR03} Sikorski, A. and Romiszowski, P.
(2003) {\it Biopolymers} {\bf 69}, 391--398.

\bibitem{KGPS03} Kolinski, A., Gront, D., Pokarowski, P., 
and Skolnick, J. (2003)
{\it Biopolymers} {\bf 69}, 399--405.

\bibitem{LHH03} Lin, C.Y., Hu, C.K., and Hansmann, U.H.E. (2003)
{\it Proteins} {\bf 52}, 436--445.

\bibitem{LMMO03} La Penna, G., Mitsutake, A., Masuya, M., and 
Okamoto, Y. (2003)
{\it Chem. Phys. Lett.}, in press.


\bibitem{FD} Falcioni, M. and Deem, M.W. (1999)
{\it J. Chem. Phys.} {\bf 110}, 1754--1766. 

\bibitem{YP} Yan, Q. and de Pablo, J.J. (1999)
{\it J. Chem. Phys.} {\bf 111}, 9509--9516. 

\bibitem{NOSMO} Nishikawa, T., Ohtsuka, H., Sugita, Y., 
Mikami, M., and Okamoto, Y. (2000)
{\it Prog. Theor. Phys. (Suppl.)} {\bf 138}, 270--271.

\bibitem{Yama} Yamamoto, R. and Kob, W. (2000)
{\it Phys. Rev. E} {\bf 61}, 5473--5476. 

\bibitem{Freeman} Calvo, F., Neirotti, J.P., Freeman, D.L., and Doll, J.D.
(2000) {\it J. Chem. Phys.} {\bf 112}, 10350--10357.

\bibitem{K02} Kofke, D.A. (2002)
{\it J. Chem. Phys.} {\bf 117}, 6911--6914. 

\bibitem{OKOM} Okabe, T., Kawata, M., Okamoto, Y., and Mikami, M. (2001)
{\it Chem. Phys. Lett.} {\bf 335}, 435--439.
\bibitem{ISNO} Ishikawa, Y., Sugita, Y., Nishikawa, T., and Okamoto, Y. 
(2001)
{\it Chem. Phys. Lett.} {\bf 333}, 199--206.

\bibitem{Gar} Garcia, A.E. and Sanbonmatsu, K.Y. (2001)
{\it Proteins} {\bf 42}, 345--354.

\bibitem{ZBG01} Zhou, R.H., Berne, B.J., and Germain, R.
(2001)
{\it Proc. Natl. Acad. Sci. U.S.A.} {\bf 98}, 14931--14936.


\bibitem{GS02} Garcia, A.E. and Sanbonmatsu, K.Y. (2002)
{\it Proc. Natl. Acad. Sci. U.S.A.} {\bf 99}, 2782--2787.

\bibitem{ZB02} Zhou, R.H. and Berne, B.J. (2002)
{\it Proc. Natl. Acad. Sci. U.S.A.} {\bf 99}, 12777--12782.

\bibitem{FMB03} Feig, M., MacKerell, A.D., and Brooks, C.L. III
(2003) {\it J. Phys. Chem. B} {\bf 107}, 2831--2836.

\bibitem{RH03} Rhee, Y.M. and Pande, V.S. (2003)
{\it Biophys. J.} {\bf 84}, 775--786.

\bibitem{GG03} Gnanakaran, S. and Garcia, A.E. (2003)
{\it Biophys. J.} {\bf 84}, 1548--1562.

\bibitem{KB03} Karanicolas, J. and Brooks, C.L. III
(2003) {\it Proc. Natl. Acad. Sci. U.S.A.} {\bf 100}, 3954--3959.

\bibitem{PS03} Pitera, J.W. and Swope, W.
(2003) {\it Proc. Natl. Acad. Sci. U.S.A.} {\bf 100}, 7587--7592.

\bibitem{FE03} Fenwick, M.K. and Escobedo, F.A.
(2003) {\it Biopolymers} {\bf 68}, 160--177.

\bibitem{SRNP03} Sorin, E.J., Rhee, Y.M., Nakatani, B.J., 
and Pande, V.S. (2003)
{\it Biophys. J.} {\bf 85}, 790--803.


\bibitem{XB00} Xu, H.F. and Berne, B.J. (2000)
{\it J. Chem. Phys.} {\bf 112}, 2701--2708. 
\bibitem{FYD02} Faller, R., Yan, Q., and de Pablo, J.J. (2002)
{\it J. Chem. Phys.} {\bf 116}, 5419--5423.

\bibitem{STREM} Mitsutake, A. and Okamoto, Y., in preparation.

\bibitem{Huk2} Hukushima, K. (1999)
{\it Phys. Rev. E} {\bf 60}, 3606--3614. 
%\bibitem{Dunw} Bunker, A. and D{\" u}nweg, B. (2000)
%\lq\lq Parallel excluded volume tempering for polymer melts,\rq\rq~
%{\it Phys. Rev. E}, in press.
\bibitem{WBS02} Whitfield, T.W., Bu, L., and Straub, J.E. (2002) 
{\it Physica A} {\bf 305}, 157--171.


\bibitem{Nose} Nos{\'e}, S. (1984)
%A molecular dynamics method for
%simulations in the canonical ensemble. 
{\it Mol. Phys.} {\bf 52}, 255--268.
%{\it Mol. Phys.} 1984, {\bf 52}, 255--268.

\bibitem{Hoover} Hoover, W.G. (1985) 
%Canonical dynamics - equilibrium phase-space distributions. 
{\it Phys. Rev. A} {\bf 31}, 1695--1697. 
%{\it Phys. Rev. A} 1985, {\bf 31}, 1695-1697.  
 

\bibitem{HLM} Hoover, W.G., Ladd, A.J.C., and Moran, B. (1982)
{\it Phys. Rev. Lett.} {\bf 48} 1818--1820.
%{\it Phys. Rev. Lett.} {\bf 48} (1982) 1818.
\bibitem{EM} Evans, D.J. and Morris, G.P. (1983)
{\it Phys. Lett.} {\bf A98}, 433--436.
%{\it Phys. Lett.} {\bf A98} (1983) 433.

\bibitem{BergTXT} Berg, B.A. {\it Markov Chain Monte Carlo Simulations
and Their Statistical Analysis I and II}, books in preparation.

\bibitem{BergLog} Berg, B.A. (2002)
%{\it Multicanonical simulations step by step}, 
cond-mat/0206333.

\bibitem{NSMO} Nagasima, T., Sugita, Y., Mitsutake, A., and
Okamoto, Y., in preparation.

\bibitem{NSMO02} Nagasima, T., Sugita, Y., Mitsutake, A., and
Okamoto, Y. (2002) {\it Comp. Phys. Commun.} {\bf 146},
69--76.

\bibitem{O03} Okamoto, Y. (2003) to appear in the 
{\it Proceedings of the Los Alamos Workshop,
The Monte Carlo Method in the Physical Sciences:  Celebrating the
50th Anniversary of the Metropolis 
Algorithm}; cond-mat/0308119.


\end{thebibliography}
\end{document}